\documentclass[12pt,preprint,epsf]{aastex} 
\pdfoutput=1

\shorttitle{Six Open Clusters }
\shortauthors{Krisciunas, Monteiro, \& Dias} 

\begin{document}
\received{30 November 2014}

\title{CCD Photometry of NGC 2482 and Five Previously Unobserved Open Star Clusters}   
\author{
Kevin Krisciunas,\altaffilmark{1}
Hektor Monteiro,\altaffilmark{2}
and Wilton Dias\altaffilmark{2}
}
\altaffiltext{1}{George P. and Cynthia Woods Mitchell Institute for Fundamental 
Physics \& Astronomy, Texas A. \& M. University, Department of Physics \& Astronomy,
  4242 TAMU, College Station, TX 77843; {krisciunas@physics.tamu.edu} }

\altaffiltext{2}{UNIFEI, IFQ - Instituto de F\'{i}sica e Qu\'{i}mica, 
Universidad Federal de Itajub\'{a}, Itajub\'{a} MG, Brazil; 
{hektor.monteiro@gmail.com}, {wiltonsdias@yahoo.com.br} }

\begin{abstract} 
We present $BV$ and
$u^{\prime}g^{\prime}r^{\prime}i^{\prime}$ CCD photometry of the
central region of NGC 2482.  We also present
$BVu^{\prime}g^{\prime}$ CCD photometry of five clusters that 
have been poorly studied in the past: Ruprecht 42, Ruprecht 51,
Ruprecht 153, Ruprecht 154; and AH03 J0748$-$26.9, which to
our knowledge has not been studied before.  Using a global
optimization technique that eliminates much of the subjectivity
previously inherent in main sequence fitting studies, we obtain
values of the distances, ages, and metallicities of the clusters,
with robust estimates of the uncertainties of these fundamental
parameters.  Four of our
clusters are less than $\sim$1.3 kpc beyond the Sun's
distance from the Galactic Center and have essentially solar
metallicity.  The metallicities of those clusters more distant from
the Galactic Center are consistent with a 0.3 dex step to lower
[Fe/H] found in other studies.
\end{abstract}


\section{Introduction}



One of the fundamental distance determination methods within the Milky
Way Galaxy is the method of main sequence fitting.  After several
decades of work by hundreds (or thousands) of astronomers, we believe
we know the absolute magnitudes of single stars of a given age and
metallicity.  Since some stars in clusters are actually unresolved
binaries, we can account for a ``leakage'' of stars in a color-magnitude
diagram (CMD) to brighter values.  
Still, if we know which objects are bona fide members
of a cluster, there should be a sharper blue edge to the main
sequence.  Blue stragglers and stars along the same line of sight
at other distances contribute to some raggedness of the blue
edge. The bottom line is that we can determine the distances of
clusters relative to the Hyades, whose distance we know from the
moving cluster method \citep{Han75} or, better still, from HIPPARCOS
parallaxes \citep{Per_etal98}.  Thus, we can determine the distances
to the clusters using broad band photometry.  As always, reddening can
be problematic.

It is our assumption that stars in a cluster are born at about the same time
and all have about the same composition.  We are therefore intrigued
by the recent finding that some {\em globular} clusters show evidence
of two, or even three, main sequences \citep{Mil_etal12, Pio_etal07}.
To our knowledge there is no known open cluster that shows clear
evidence of multiple stages of star formation.

Since the publication of the catalogue of open clusters of \citet{Alt_etal58}
and its Supplements, the number of clusters has risen from 945 to more than 2000
\citep {Dia_etal02}.\footnote[3]{The Dias catalogue is available online at
http://www.astro.iag.usp.br/ocdb/.} \citet{Oli_etal13} point out that
fewer than 10 percent of 2174 catalogued open clusters have their metallicity
determined in the literature.

In this paper we present Johnson $BV$ photometry and photometry using filters of 
the Sloan Digital Sky Survey \citep{Fuk_etal96}.  We have selected six clusters 
that have been poorly studied and one that, to our knowledge, has no published 
optical or near-IR photometry. We apply to the data the cross-entropy technique, 
first introduced by \citet{Rub99}, with the objective of estimating probabilities 
of rare events in complex stochastic networks, as previously done by 
\citet{Mon_etal10}, \citet{Oli_etal13}, and references therein.  The analysis 
uses a weighted likelihood criterion to define the goodness of fit and global 
optimization (i.e. cross-entropy) to find the best fit isochrones. For each 
cluster we derive the distance, age, $B-V$ color excess, and metallicity.  For 
four clusters with moderate color excess we also derive R$_V$ = A$_V$/E($B-V$), 
which we set as a free parameter.

The clusters studied make a relevant contribution towards the
establishement of a large sample of objects that have fundamental
parameters, if not the observational data itself, determined in a
homogenous way. Such a sample can be important in studies of Galactic
structure and evolution.


In this paper we adopt a solar metallicity of Z = 0.019, the value
adopted for the set of isochrones of \citet{Mar_etal08}, which we use.
The isochrone set of \citet{Mar_etal08} was chosen because of its
widespread use in the literature as well as its ease of use and
implementation in our code. We refer the reader to that work for
details and dicussion of the assumptions made by the authors. The
distance of the Sun from the Galactic Center is taken to be 8.0 kpc
\citep{Reid93}.

\section {Data Acquisition and Reduction}

\subsection {The Clusters}

Basic information for the clusters observed is given in Table \ref{clusters}.
Details of the images used for this paper are given in 
Table \ref{exptimes}, in the Appendix.  The Appendix also gives transformation
equations from instrumental magnitudes and colors to standardized values, and 
the reduction coefficients assumed and derived.

NGC 2482 (originally designated H VII 10) was discovered by
William Herschel on 20 November 1784 with his 19-inch diameter reflector
of 20-ft focal length \citep[][on p. 496]{Her1786}.\footnote[4]{
A further examination of the observational record (Steinicke 2013,
private communication) indicates that William
Herschel saw the cluster again on 3 June 1785.  His son John Herschel saw it
with his own 18.25-inch reflector of 20-ft focal length on 7 January 1831
(catalogued as h 474).  John Herschel saw it again on 28 January 1837 at the Cape
of Good Hope with the same telescope (catalogued as h 3106).}  He described
it as, ``A very large cluster of scattered stars extremely
rich and compressed, more than 15$\arcmin$ in diameter.''

According to \citet{Dia_etal02}
NGC 2482 has an {\em apparent} distance modulus of 10.93 mag (meaning
that extinction has not been taken into account), a distance
of 1343 pc, a color excess E($B-V$) = 0.093 mag, an age of
400 Myr, and an iron abundance of [Fe/H] = +0.12 dex.  This abundance
value is based on CMT$_1$T$_2$ Washington broad-band photometry
of three giant stars by \citet{ClaLap83}.  More recently, 
\citet{Red_etal13} determined
the metallicity of one of those stars, star 9 in the list of
\citet[][Table 18]{MofVog75}, to be [Fe/H] = $-$0.07 $\pm$ 0.04.

\citet{MofVog75} provide $V$ magnitudes, $B-V$ and $U-B$ colors
of 41 stars in the direction of NGC 2482.  Seventeen of the
stars are designated non-members of the cluster.  Five of the 41 
stars are considerably to the west (up to 15$\arcmin$) of 
the center of the cluster, near BX Puppis; one of these has 
photometry consistent with it being a cluster member.  Most 
of the stars are distributed over a triangular region about 
14$\arcmin$ by 11$\arcmin$ by 12$\arcmin$ in angular size.

According to \citet{Twa_etal97}, ``The [color-magnitude diagram]
of \citet{MofVog75} is too ill-defined to permit a reliable distance
determination.  Using the DDO modulus adjusted for the Hyades
distance, one gets a very uncertain ($m-M$) = 10.3.''  This corresponds
to a distance of 1148 pc.

To our knowledge the only reference to AH03 J0748$-$26.9 is in the
catalogue of clusters by \citet{ArcHyn03}.  Ruprecht 42, 51, 153 and
154 are listed by \citet{Rup66}.  References cited in that work
provide more information.  Prior to our observations and analysis
presented here, we were aware of no published photographic photometry,
photoelectric photometry, or CCD photometry of these five 
clusters.\footnote[5]{However, see comments below concerning the
catalogue of \citet{Kha_etal13}.}

\subsection{Details of the Imagery}

The data presented in this paper were obtained by one of us (KK) at Las 
Campanas Observatory (LCO) using the 1-m Henrietta Swope telescope.  Three 
second exposures of NGC 2482 were obtained in $BVgri$ on 6 January 2012 (UT), 
along with a 24 second $u$-band exposure.  Longer $BV$ exposures of 150 and 
120 seconds, respectively, were taken on 21 December 2012 (UT).  For the NGC 
2482 images we used the central 1201 $\times$ 1201 pixels of the CCD chip then 
in use. The pixel scale is 0.435 arc seconds; the field size is 8.7 by 8.7 arc 
minutes.

The other clusters were observed from 31 December 2013 to 4 January 2014 
(UT) using a new CCD camera having four 2K by 2K chips.  We observed all 
targets with chip number 3, offsetting 300 arcsec north and east of the 
center of the four chip array.  We found after the fact that the point 
spread function (PSF) is good for chip 3 for X and Y ranging from about 600 
to 2048 pixels.  Along the left and bottom of chip 3, furthest from the 
center of the camera field of view, the PSF contours can significantly 
deviate from circularity. This is particularly the case for $B$-band images.  
We derived positions and magnitudes for stars over the whole of chip 3, but 
for the analysis of the clusters we restricted ourselves to a subset of the 
area covered.  Given the angular size of the clusters, this feature of the 
PSF actually did not come into play in a significant way.  The pixel scale 
is the same, so these images are 14.8 by 14.8 arc minutes in size.

For these five clusters we typically took one exposure per filter at or near 
the limit of the shortest exposure possible (5 seconds), and one or two 
exposures that were much longer (up to 120 sec).  This minimizes the number 
of saturated stars and expands the dynamic range of the photometry.

The data presented here were all obtained on photometric nights, though on
some occasions the seeing was worse at the start of the night.  We observed
standards of \citet{Lan92} and \citet{Smi_etal02} on five or six occasions
each night.   

\parindent = 9 mm

\subsection{The Photometry}

Not only is NGC 2482 a good candidate for modern
analysis, but its brightest stars did not reach the non-linear
response level or saturation level of the shortest exposures
possible with the CCD camera installed on the Las Campanas 1-m Swope 
telescope through 2012.  This cluster was previously studied
by \citet{MofVog75} using single channel photoelectric photometry.
Our PSF photometry, obtained from images of 6 January 2012,
is given in Table \ref{photometry}.  We give $V$ and $r^{\prime}$
magnitudes and four photometric colors.

The coordinates of the NGC 2482 stars in Table \ref{photometry} were 
determined using {\sc ccmap} and {\sc cctran} in the {\sc imcoords} 
package of {\sc iraf}.\footnote[6]{{\sc iraf} is distributed by the 
National Optical Astronomy Observatory, which is operated by the 
Association of Universities for Research in Astronomy, Inc., under 
cooperative agreement with the National Science Foundation (NSF).}  
We used the X-Y pixel coordinates of 10 stars distributed over the 
whole field and the right ascensions and declinations from a 
Digital Sky Survey\footnote[7]{http://archive.stsci.edu/cgi-bin/dss\_form} 
image displayed with DS9 to obtain the transformation solution.


In Table \ref{ids} we show the correspondence of our ID's and those of
\citet{MofVog75}. There are 17 single stars in common.  
With ``$\Delta$'' in the sense ``our aperture photometry {\em minus} that of
\citet{MofVog75}'', we find $\langle \Delta$ ($V$)$\rangle$ = +0.024 mag, 
with a standard deviation of the distribution of $\pm$ 0.035.  We find 
$\langle \Delta$ ($B-V$)$\rangle$ = +0.001 mag, with a standard deviation
of the distribution of $\pm$ 0.024.  The average difference of our
estimated $U-B$ colors compared to those of \citet{MofVog75} is 
$-$0.033 mag, with a standard deviation of the distribution of
$\pm$ 0.056.  Thus, within the errors, our photometry of the brighter stars
is in statistical agreement with previously published values.

We note that two of the stars of \citet{MofVog75}  show close 
companions in our imagery. A combination of the photometry of
our stars 241 and 239 is in reasonable agreement (within 0.04 mag) with 
their star 17. Combined photometry of our stars 220 and 218 is in reasonable 
agreement with data of their star number 20.

In Fig. \ref{ngc2482_rgr} we show an $r^{\prime}$ vs. $g^{\prime} - 
r^{\prime}$ color magnitude diagram of NGC 2482.  The $V$ vs. $B-V$ CMD is 
the middle two panels of Fig. \ref{ngc2482}. 11 of our 114 stars are obvious 
non-cluster members.  They are presumably red giants at other distances.  
The brightest star in the field (our star 166) is a bona fide giant in the 
cluster.

As noted above, the other five clusters were observed with a
new CCD camera using a larger chip, but with the same plate scale.
For these clusters we obtained photometry of a total of 3871 stars 
that satisfied the S/N criteria.  The coordinates of the stars
(in decimal degrees, and pixel coordinates) and the derived
$UBV$ photometry can be obtained by the reader via this link: 
people.physics.tamu/krisciunas/clusters.tar, which also contains
data for 114 stars of NGC 2482 derived from PSF photometry.  In the final
column of each of these files we flag the stars most likely to be
cluster members (normalized probability greater than 50 
percent).\footnote[8]{We note that the normalized probabilities are
not membership probabilities in the classical sense, from proper
motion vectors and/or radial velocities.  The cross-entropy method
maximizes a likelihood measure for an {\em ensemble} of stars.
Even stars with low likelihood as members contribute to the multi-parameter
solution.}  The coordinates given in the files
are basically for the purpose of identification, not for astrometric
studies requiring extreme accuracy.

\subsection {Aperture Photometry vs. PSF Photometry}

It is customary in star cluster studies to rely primarily on PSF 
photometry. This makes sense, as most open clusters are found at 
low Galactic latitude. There can be thousands of stars in the 
field, resulting in severe challenges for aperture photometry.  To 
obtain the instrumental PSF magnitudes of our stars, we used a 
stand alone version of {\sc daophot} \citep{Ste87} and {\sc 
allstar}.\footnote[9]{We found useful a web document called ``A 
primer for DAOPHOT II and ALLSTAR'' by Bill Harris (April 2002) and 
further extended by Helmut Jerjen.  See: 
www.mso.anu.au/$\sim$jerjen/daophot.html} It is straightforward to 
keep track of the pixel coordinates and instrumental magnitudes, 
and adjust those magnitudes to the equivalent values for a 1.0 
second exposure.  However, instead of using {\sc daomatch} and {\sc 
daomaster}, we wrote our own {\sc fortran} software to 
cross-correlate the frames taken in different filters.  We used the 
following selection criteria.  We were interested in stars observed 
in {\em all} filters, 18-$\sigma$ (or more) above the sky level in 
the $u$-band, and 30-$\sigma$ (or more) above the sky level in the 
other filters.  These selection criteria had the effect of 
eliminating many stars that are very red in $u^{\prime} - 
g^{\prime}$ and much redder than the main sequence of each cluster 
studied.

We should mention that one particular advantage of PSF photometry is
that we can exploit our knowledge of the PSF to obtain a higher signal
to noise ratio (S/N) for faint stars.  We can dig deeper into the
image.  This is not too relevant to this paper however, since we
wanted photometry accurate to 0.06 mag or better,
so we limited ourselves to relatively bright stars, where the turnoff
point of an open cluster typically lies.

Aperture photometry has its own advantages.  Providing that the software
aperture radius is sufficiently large (for the LCO 1-m we find this to
be typically 8 pixels), accurate photometry can be ensured on photometric
nights even if the PSF is considerably non-circular.  The {\sc iraf}
package {\sc apphot} subtracts the sky level using a settable annulus,
and if one uses the {\em median} counts in that annulus, cosmic ray hits
or the presence of other stars in the annulus have no effect on the
results. We typically used a sky annulus ranging from 12 to 20 pixels.
Thus, aperture photometry {\em can} be used in moderately
crowded fields.  

Using the {\sc ccdred} package in {\sc iraf} we bias-corrected our
images using a nightly master bias frame, flattened and trimmed them.
The quality of each night was evaluated with observations of standards
of \citet{Lan92} and \citet{Smi_etal02}. For this we used the {\sc
photcal} package in {\sc iraf}.  This gave us photometric zero
points and color terms.  (See Appendix A and Table \ref{transform_params}.)
If the standards were
observed over a sufficiently wide range of air mass, we also derived
the atmospheric extinction and reddening terms.  We did this for each
night using the instrumental magnitudes of the standards derived from
aperture photometry {\em and} from PSF photometry.  Not surprisingly,
the color terms and atmospheric terms were the same, within the
errors, using the two methods.  The instrumental magnitudes from PSF
photometry (adjusted to a 1 second exposure time) are typically 0.02
to 0.04 mag brighter than the instrumental magnitudes from aperture
photometry using an 8 pixel radius aperture.  The PSF magnitudes include
light most or all of the wings of the profile.  And so the photometric
zero points for transformations like Eqn. 1 are different by such an
amount.  One would have to use a very large software aperture to
include ``all'' of the flux for the aperture magnitudes, and this
greatly increases the chances of including other stars in the
aperture.  Our motivation here was to have the most robust reduction
parameters to convert the PSF magnitudes and colors to standardized
values using simple {\sc awk} scripts operating on our text files of
raw PSF data.

In Fig. \ref{finder} we show a finder chart of 114 stars of NGC 2482 that
were brighter than 18-$\sigma$ above the sky level in the $u$-band, 
and 30-$\sigma$ above the sky level in the other filters.  Given the
field size in pixels, this amounts to one ``bright'' star, on average, 
for every 112 by 112 pixel square of the image.  Surely, 
by avoiding obvious close pairs, the results of aperture photometry and 
PSF photometry should be comparable.


A comparison of photometry done with the two methods is shown in
Fig. \ref{pm_comp}. Aperture photometry of stars fainter than $V \approx$ 13.2
was obtained on 21 December 2012. All PSF photometry comes from 6 January 2012.
We have also accounted for a systematic error of 78 msec in the shutter time 
of the older camera at the LCO 1-m telescope \citep[][Appendix A]{Ham_etal06}.  

Fig. \ref{pm_comp} shows that there are no significant trends in the differences
of aperture and PSF photometry of NGC 2482 down to $V \approx$ 15.7.
The scatter of the individual differences increases, as expected, at fainter
magnitudes, but the data are consistent to 0.06 mag or better.

We note that if one's ultimate goal is to produce color-magnitude
diagrams, the internal errors are smaller for the colors if one sets
up the data reduction to produce one magnitude and multiple colors
using transformations specified in the configuration file (e.g in {\sc
  fitparams} in {\sc iraf}).  In our experience at Cerro Tololo and
LCO we find that the RMS scatter of the photometry of the standards on
a photometric night is typically $\pm$0.025 mag in $V$, and $\pm$0.015
to 0.020 mag for $B-V$ colors. (These uncertainties should be added in
quadrature to the internal random errors of the program objects.)  If
one configures {\sc fitparams} and {\sc evalfit} to produce $B$-band
magnitudes directly, the RMS error might be $\pm$0.025 mag or more.
If one {\em then} obtains the $B-V$ colors by subtracting $V$
magnitudes from $B$ magnitudes, simple statistics stipulates that we
would obtain uncertainties of $\pm$0.035 mag or more, roughly twice
the RMS scatter for the colors obtained the other way using the same
instrumental magnitudes. This is not a critical issue for
stellar photometry, but it is something one should consider.

\section{Analysis and Discussion}

The cross-entropy technique applied to the isochrone fits done in
this work is described in detail in a series of papers.  See
\citet{Oli_etal13} and references therein. It eliminates much of the
subjectivity inherent in open cluster fitting.  Using $UBV$
photometry, it simultaneously solves for the reddening, distance, age,
and metallicity of a cluster using an estimation of membership
likelihood as described by \citet{dias_etal12}.  A new feature of
the program implemented in this work is that if the
color excess of a cluster is greater than some settable value
(we chose 0.1 mag), a value of R$_V$ =
A$_V$/E($B-V$) is also derived.  Otherwise, R$_V$ = 3.1 is assumed.
The multi-parameter space considered
is as follows: 1) log (age in years) from 6.60 to 10.15 in steps of
0.05; 2) 1 to 10,000 pc in distance; 3) E($B-V$) from 0.0 to 3.0 mag;
4) metallicity from 0.0001 to 0.03 in steps of 0.05 in log Z; and 5)
R$_V$ from 2.0 to 6.0. From a detailed analysis of nine well studied 
open clusters \citet{Oli_etal13} show that their technique works well.

Since the determination of the age of the cluster depends most
critically on the hottest main sequence stars, those just past the
turnoff point, and the red giants of the cluster, the lower main
sequence stars are not critical but do help in constraining the
general shape of the main sequence, especially with the $U$-band
data.  As a result, we may either set a limit on the faintest stars
to consider, or let the program eliminate from consideration the stars
fainter than the maximum of a histogram of the $V$-band magnitudes, so
as to have a sample reasonably complete to some brightness level.  An
extra piece of information we have on each star is its angular
distance from the cluster center.  If the images are sufficiently big
to cover the whole extent of the cluster, the program can derive the
most likely location of the center from the peak of a density map
obtained with a Gaussian kernel of predefined width.  One of the
few arbitrary (but settable) parameters is the assumed fraction of
unresolved binaries.  With one exception we adopted 50 percent for
this fraction. For NGC 2482 we adopted 40 percent for this fraction,
as we found that it slightly improved the likelihood measure of our
solution.  Then, using a \citet{Sal55} mass function, we performed Monte
Carlo simulations to pick at random the particulars of the companions.
Carrying out a number of runs (typically 50) gives us the
uncertainties of our derived parameters.  Which stars are assigned
binary status varies from run to run.  Finally, in crowded fields with
many giant star interlopers, some additional cuts by hand 
might be necessary on the basis of angular distance from the 
cluster center, the colors of the stars in the field and other data.

Concerning the reality of our targets being bona fide clusters, two of
our targets (Ruprecht 153 and 154) were problematic, owing in part to
the distribution of non-cluster stars in the field. This is a
limitation of our method.  For very crowded fields our procedure
may fail to assign low weights to non-member stars 
solely on the basis of photometric data. 
In Fig. \ref{density}
we show a comparison of the density maps of the six clusters.
For Ruprecht 153 we used a Gaussian kernel with 
$\sigma= 2.5\arcmin$. For the other clusters we used a kernal
of 1\arcmin.
 
For Ruprecht 154 we ran 100 iterations of the bootstrap procedure in
order to sample better the distribution of solutions. The results show
a clear double peaked distribution for the ages and non-Gaussian
distributions for distance and reddening. The results are shown in
Fig. \ref{rup154-hists}, where we also present a 2D histogram in age
versus distance showing that in this parameter space there are two
well defined solutions with distinct ages and slightly different
distances. The solution of lower age and distance seems to be the most
probable, which is what we adopt. However, the results can be
alternatively interpreted as two superimposed populations.  Due to the
small number of stars that have high likelihood of being cluster
members, the results seem to be inconclusive for Ruprecht 154, since
we can not rule out the possibility of chance alignment in the plane
of the sky. A visual inspection of the proper motion vectors for
the fields of both Ruprect 153 and Ruprecht 154 seem consistent with
the latter hypothesis. Given all these considerations we believe
that Ruprecht 153 and 154 warrant further study, including
detailed analysis using proper motions, to better define
their fundamental parameters or even their open cluster status.
This, however, is beyond the scope of the present paper.


Results for the clusters observed are to be found in Table \ref{results}. 
The six clusters studied have distances ranging from 1194 pc (NGC 2482) to 5457 
pc (Ruprecht 42).  The range in age is from 63 Myr (AH03 J0748$-$26.9) to 
2.2 Gyr (Ruprecht 154). The weighted mean value of R$_V$ for four clusters 
with moderate reddening is 3.06 $\pm$ 0.22, typical for ``normal'' Milky Way 
dust with R$_V$ = 3.1 \citep{Car_etal89}.  In the final column of Table
\ref{results} we give the usual metallicity parameter [Fe/H] = log (Z/0.019).

Color-color diagrams and color-magnitude diagrams of the five clusters
observed in December, 2013, and January, 2014, are shown in Figs.
\ref{ah03} through \ref{rup154}.  

We have run the cross-entropy code on the PSF photometry {\em and} the 
aperture photometry of NGC 2482.\footnote[10]{To provide a robust test 
of the consistency of the two data reduction methods, we ran the 
cross-entropy code on the {\em exact same} sample of 92 stars.  Those 
identified by {\sc daophot} included some stars close to the edges of 
the chip and obvious close pairs.  Many stars selected by eye for 
aperture photometry from the longer and shorter $V$-band exposures are 
quite red and failed the 18-$\sigma$ S/N criterion in the $u$-band for 
PSF photometry.}  Using PSF photometry we obtain an age of 447 Myr, a 
distance of 1194 pc, E($B-V$) = 0.094, and Z = 0.015. Using aperture 
photometry we obtain an age of 371 Myr, a distance of 1349 pc, E($B-V$) 
= 0.086, and Z = 0.022.  Given the uncertainties of the results given 
in Table \ref{results}, the corresponding differences are not 
statistically significant. The reader is free to average our two solutions
for this cluster.  
We note that our value of [Fe/H] = $-$0.10$^{+0.13} _{-0.18}$ from PSF 
photometry is in excellent agreement with the value $-0.07 \pm 0.04$ 
from spectroscopy of one giant in the cluster by \citet{Red_etal13}.

It is known that there is a decrease of metallicity with increasing distance 
from the Galactic Center.  Many authors, such as \citet{RyuLee11}, fit a 
straight line to the data, obtaining a gradient of [Fe/H] amounting to 
$-$0.076 dex per kpc. Other authors show that it is more sensible to adopt a 
step function in the metallicity \citep{Twa_etal97, Lep_etal11}.  The 
metallicity is essentially constant, and comparable to that of the Sun, out 
to $\sim$1 kpc beyond the Sun's distance from the Galactic Center, then drops to 
[Fe/H] $\approx -0.30$ beyond that. In Fig. \ref{metals} we show that our 
results are consistent with such a step function.

In Fig. \ref{dias_cat} we present a plot of the positions of several
hundred clusters from the catalogue of \citet{Dia_etal02}
(those with the X-Y range shown), and our six 
clusters.  We have coded the points by colors corresponding to the range of 
age in log $t$ (in years).  All of our clusters are situated in the Orion spur 
of the Galaxy, or in the Perseus arm; see Fig. 16 of \citet{Chu_etal09}.
We note that of 1985 clusters in the catalogue of \citet{Dia_etal02} with
ages specified, 80.6 percent have Galactic Z distances between $-$200 and +150 pc
with respect to the Sun.  There are only 21 clusters with $\mid$Z$\mid >$ 1000 pc
in the catalogue.  The complex picture shown in Fig. \ref{dias_cat} arises
in part because we did not carry out any analysis of subsets of the Dias
catalogue according to Galactic Cartesian coordinates.

Since we are interested in open clusters for distance determinations, we note 
that for our six clusters we obtain a mean {\em relative} uncertainty of 
distance of 8.8 percent.  This corresponds to an uncertainty in the distance 
modulus of $\pm$0.183 mag. This may be compared to $\sigma _V$ = $\pm$0.159 
mag from 649 fundamental mode Cepheids in the Large Magellanic Cloud and 
$\sigma _V$ = $\pm$0.257 mag from 466 fundamental mode Cepheids in the Small 
Magellanic Cloud \citep[][Table 1] {Uda_etal99}.  The intrinsic scatter of the 
absolute magnitudes of RR Lyrae stars is $\pm$0.24 mag or less 
\citep{Lay_etal96}. Type Ia supernovae at maximum light in the near-infrared 
are almost perfect standard candles.  We know their absolute magnitudes to 
$\pm$0.15 mag or better \citep{Kri12}.

It is also important to make a few points about the error estimates in
this work, especially regarding the accuracy and precision of the
method. It was shown by \citet{Mon_etal10} that the method is accurate
when evaluated against a synthetic cluster so that we have complete
knowledge of the parameters used. This of course is not the case for
real clusters where we have no ``true'' value to compare the fit
results to. In this context our error estimates can be said to be
precise, given the size of the errors relative to the parameters
determined. However, this is completely dependent on the
characteristics of the data used. So for a dataset that has many
stars with low contamination, or contamination that can be removed
easily, we will have very precise results, that is, with small
error. The errors we quote are related to the precision of the {\em fit}. 
In other words, for a given dataset and likelihood
function, weights, and isochrone grid assumptions used, the errors quoted
are the standard deviations of the distribution of solutions obtained
from the bootstrap procedure.

As we were finishing this paper we learned of the catalogue of 3006 
bona fide objects of \citet{Kha_etal13}, which contains entries for 
five of our six clusters: NGC 2482, Ruprecht 42, 51, 153, and 154. 
Their parameters for NGC 2482 are in agreement with our results
thanks to previously published optical data by \citet{MofVog75},
but their parameters for the four Ruprecht clusters differ 
considerably from ours. For example, their entry for 
Ruprecht 154 gives a color excess E($B-V$) = 0.781 mag and an age 
of 14 Myr, while we obtain 0.087 mag and 2.2 Gyr, respectively, for 
the same cluster!  Also, note how different is the isochrone 
corresponding to their value of the cluster age of Ruprecht 42 in 
Fig. \ref{rup42}.  

\citet{Kha_etal13} used 
PPMXL,\footnote[11]{http://irsa.ipac.caltech.edu/Missions/ppmxl.html}
a catalog of positions, proper motions, optical photometry, and
near-IR photometry from the Two Micron All-sky
Survey.\footnote[12] 
{http://www.ipac.caltech.edu/2mass/releases/allsky/index.html; see 
also \citet{Skr_etal06}.}  While \citet{dias_etal12} have demonstrated
that 2MASS photometry alone {\em can} lead to sensible results
for some clusters, we find that a near-IR color-magnitude of 
2MASS data of AH03 J0748$-$26.9 reveals no obvious main sequence,
while our optical data clearly shows one.  An attempt by us to
fit the near-IR data of Ruprecht 154 also yielded no scientific fruit.
We surmise that the photometric depth 
and accuracy of the 2MASS catalogue alone is not enough to resolve 
the main sequence and turnoff point of many clusters superimposed
on crowded fields.  Since \citet{Kha_etal13} have based some of
their catalogue entries on 2MASS data, without the use of optical
photometry, the reader should treat some of the individual 
entries of their catalogue with caution.


\section{Conclusions}

We have presented broad band photometry of six Galactic open clusters,
four of which were poorly studied (no optical CCD photometry) and
one that has never been studied before.  Using the cross-entropy
method applied to the fitting of theoretical isochrones to
photometric data as detailed by \citet{Mon_etal10}, for each
cluster we have derived the age, distance, reddening, and metallicity,
with robust error bars.

Four of our six clusters, those less than 1.3 kpc beyond the Sun's
distance from the Galactic center, have metallicity slightly greater
than that of the Sun
($\langle$Z$\rangle$ = 0.022 $\pm$ 0.002), while two of our clusters
beyond this distance have lower than solar metallicity
($\langle$Z$\rangle$ = 0.012 $\pm$ 0.003).  The individual values,
converted to [Fe/H] in the usual way, are consistent with the step
function found by \citet{Twa_etal97}, and \citet{Lep_etal11}.

As mentioned by \citet{Oli_etal13}, only about 10 percent
of 2174 catalogued open clusters have published values of metallicity.
Many of these are obtained from spectroscopy, which is the preferred
method. However, observationaly, this is a costly endeavor and
alternatives that allow for the determination of the metallicity for a
large number of clusters are desirable. In this context, values
determined from photometry as done in this work can have an important
impact in the research fields that use metallicity as a tool. Efforts
to complete the catalog may lie far in the future and may require a
large scale Galactic plane survey, something that may soon be possible
using modern instruments and techniques. The tools and procedures such
as those used in this work will surely be important in that
scenario.

We suggest three worthwhile follow up endeavors: 1) Obtain high
resolution spectra of several stars brighter than $V \approx$ 14
in NGC 2482 that are $\approx$ 0.3 mag redder or bluer than the
median colors of other presumed main sequence stars of comparable brightness.  
This will shed light on the issue of unresolved binaries.
2) A much bigger project is to obtain accurate proper
motions of our six clusters in order to eliminate obvious non-cluster stars.
Since the mean proper motions of these clusters are only a few
milliarcsec per year, this will be a challenge.  3) Carry out a
detailed analysis of the spatial and kinematical structure of the
Galaxy.  This would be, in effect, a four dimensional version
of our Fig. \ref{dias_cat} (three spatial dimensions and one
of time).

\vspace {1 cm}

\acknowledgments

We thank Woody Sullivan and Wolfgang Steinicke for their intimate 
knowledge of William Herschel's observations. We thank Lucas Macri 
for installing a stand alone version of {\sc daophot} and {\sc 
allstar}, and for useful references.  Wenlong Yuan and Alejandro 
Lorenzo helped with data analysis.  We thank an anonymous referee
for useful suggestions.  We particularly thank Mark Phillips for the 
opportunity to take the imagery presented here while observing for 
the Carnegie Supernova Project.  The observations presented here 
were a backup program for times when there were no reasonably 
bright supernovae high in the sky.  

H. Monteiro would like to thank FAPEMIG for the grant CEX-PPM-00235-12
and Cnpq grant 306632/2013-6.

\appendix
\section{Further Details of the Imagery and Data Reduction}

In Table \ref{exptimes} we give the exposure times and seeing (in pixels)
of the imagery used for this project.  Table \ref{transform_params} gives
the adopted (or derived) atmospheric extinction and reddening parameters
used for the data reduction along with the derived color terms, photometric
zeropoints, and the RMS uncertainties of the nightly fits.  The camera
used for the 2012 imagery of NGC 2482 was replaced prior to the December 
2013/January 2014 observing run.  One can see that the color terms obtained
from the observations of standard stars are consistent from night to night
for observations with the same camera.

Typical transformations from the instrumental magnitudes to standardized
ones are as follows:

\begin{equation}
V \; = \; v \; - \; k_v X_v \; + \; \epsilon _V (b-v) \; + \; \zeta _V \;    ;
\end{equation}

\begin{equation}
B \; = \; b \; - \; k_b X_b \; + \; \epsilon _B (b-v) \; + \; \zeta _B \;    ;
\end{equation}


\begin{equation}
u^{\prime}-g^{\prime} \; = \; \mu _{ug} (u-g) \; - \; k_{ug} \bar{X}_{ug} \; + \; \zeta _{ug} \;  .
\end{equation}

\parindent = 0 mm

Comparable transformation equations can be written to produce $r^{\prime}$,
$g^{\prime} - r^{\prime}$ and $r^{\prime} - i^{\prime}$.
The values on the right hand sides ($u$, $b$, $v$, $g$) are instrumental magnitudes. $k$ 
values are atmospheric extinction or reddening terms. X (or $\bar{X}$) is the air mass 
value for an observation in a given filter (or the average of the air mass values for two 
filters), $\epsilon$'s and $\mu$'s are color terms, and $\zeta$'s are photometric zero 
points.  The values on the left hand sides are standardized $V$ and $B$ magnitudes 
in the \citet{Lan92} system. $u^{\prime}-g^{\prime}$ is the color in the Sloan 
``primed'' system of \citet{Smi_etal02}.

\parindent = 9 mm

A transformation of our measured $u^{\prime} - g^{\prime}$ colors to 
$U-B$ colors can be obtained from the values for the standards
observed on a particular night given by \citet{Lan92} and \citet{Smi_etal02}.
For example, given the standards used on 6 January 2012, we obtain::  

\begin{equation}
(U-B)_{Landolt} \; = \; (0.792 \pm 0.018) \; (u^{\prime} - g^{\prime}) \; - \; (0.877 \pm 0.031) \; .
\end{equation}


\clearpage

\begin{deluxetable}{lcccccccr}
\tablecolumns{9}
\tablewidth{0pc}
\tabletypesize{\scriptsize}
\tablecaption{Clusters Observed\tablenotemark{a}\label{clusters}}
\tablehead{ \colhead{Cluster} & \colhead{RA} & \colhead{DEC} & \colhead{$l$} & 
\colhead{$b$} & \colhead{$\mu _{RA}$} & \colhead{$\mu _{DEC}$} &
\colhead{Filters} & \colhead{N} }
\startdata
AH03 J0748$-$26.9   & 7:48:40 & $-$26:56:48 & 243.183 & $-$0.606 & \ldots          & \ldots          & $uBVg$   &  708 \\
NGC 2482            & 7:55:12 & $-$24:15:30 & 241.626 &   +2.035 & $-$03.29 (0.17) &   +03.29 (0.15) & $uBVgri$ &  114 \\
Ruprecht 42         & 7:57:36 & $-$25:55:00 & 243.326 &   +1.639 & $-$02.67 (0.39) &   +03.41 (0.05) & $uBVg$   &  958 \\
Ruprecht 153        & 8:00:19 & $-$30:17:00 & 247.360 & $-$0.139 & $-$02.40 (0.19) & $-$00.39 (0.35) & $uBVg$   &  586 \\
Ruprecht 154        & 8:01:46 & $-$44:25:00 & 259.581 & $-$7.302 & $-$05.16 (0.93) &   +02.67 (0.75) & $uBVg$   &  461 \\ 
Ruprecht 51         & 8:03:37 & $-$30:39:00 & 248.048 &   +0.270 & $-$02.81 (0.22) &   +00.16 (0.28) & $uBVg$   & 1158 \\
\enddata
\tablenotetext{a}{For each cluster we give the RA and DEC for equinox J2000 of 
the nominal cluster center, the corresponding Galactic longitude and latitude
(in degrees), the mean proper motion of the cluster in RA and DEC (milliarcsec per year),
the filters used, and the number of stars identified by DAOPHOT
that satisfied the apparent magnitude criteria (18-$\sigma$ over sky 
in $u$ and 30-$\sigma$ over sky in the other filters).}
\end{deluxetable}



\newpage

\begin{deluxetable}{ccccccrrl}
\tablecolumns{9}
\tablewidth{0pc}
\tabletypesize{\scriptsize}
\rotate
\tablecaption{Photometry of NCG 2482 Central Region\tablenotemark{a}\label{photometry}}
\tablehead{ \colhead{RA\tablenotemark{b}} & \colhead{DEC\tablenotemark{b}} & 
\colhead{$V$} & \colhead{$B-V$} & \colhead{$u^{\prime} - g^{\prime}$} & 
\colhead{$r^{\prime}$} & \colhead{$g^{\prime} - r^{\prime}$} & \colhead{$r^{\prime} - i^{\prime}$} & \colhead{ID} }
\startdata
  7:54:56.64 & $-$24:12:59.9 & 15.677 (0.034) &  0.476 (0.057) &  1.419  (0.078) &  15.588 (0.046) &     0.269 (0.046) &     0.065 (0.042) &  86 \\
  7:54:56.76 & $-$24:15:52.9 & 15.058 (0.031) &  0.613 (0.054) &  1.235  (0.071) &  14.926 (0.041) &     0.342 (0.041) &     0.136 (0.040) &  174 \\
  7:54:56.76 & $-$24:16:05.2 & 15.697 (0.042) &  0.511 (0.062) &  1.073  (0.072) &  15.514 (0.039) &     0.390 (0.039) &     0.151 (0.039) &  178 \\
  7:54:57.68 & $-$24:11:54.3 & 12.819 (0.029) &  0.258 (0.033) &  1.217  (0.059) &  12.848 (0.031) &  $-$0.018 (0.031) &  $-$0.066 (0.039) &  52 \\
  7:54:58.37 & $-$24:13:42.5 & 11.693 (0.024) &  0.107 (0.024) &  1.128  (0.058) &  11.774 (0.030) &  $-$0.146 (0.030) &  $-$0.173 (0.033) &  112 \\
  7:54:58.97 & $-$24:16:00.8 & 11.228 (0.024) &  0.082 (0.028) &  1.118  (0.057) &  11.308 (0.028) &  $-$0.173 (0.028) &  $-$0.177 (0.030) &  177 \\
  7:54:59.14 & $-$24:15:50.6 & 12.379 (0.024) &  0.184 (0.026) &  1.186  (0.056) &  12.435 (0.029) &  $-$0.075 (0.029) &  $-$0.129 (0.030) &  172 \\
  7:54:59.32 & $-$24:12:10.8 & 15.154 (0.038) &  0.718 (0.052) &  1.580  (0.074) &  14.900 (0.039) &     0.535 (0.039) &     0.237 (0.048) &  60 \\
  7:54:59.39 & $-$24:15:48.5 & 15.616 (0.035) &  0.366 (0.052) &  1.404  (0.068) &  15.534 (0.040) &     0.217 (0.040) &     0.038 (0.045) &  168 \\
  7:55:01.21 & $-$24:11:17.3 & 15.030 (0.040) &  0.595 (0.052) &  1.190  (0.069) &  14.871 (0.044) &     0.354 (0.044) &     0.132 (0.050) &  32 \\

  7:55:01.25 & $-$24:17:44.1 & 15.649 (0.037) &  0.443 (0.053) &  1.519  (0.077) &  15.519 (0.037) &     0.289 (0.037) &     0.068 (0.046) &  231 \\
  7:55:02.02 & $-$24:18:15.0 & 13.934 (0.027) &  0.426 (0.030) &  1.102  (0.056) &  13.847 (0.028) &     0.238 (0.028) &     0.031 (0.033) &  251 \\
  7:55:02.15 & $-$24:12:59.4 & 15.273 (0.039) &  0.371 (0.048) &  1.449  (0.071) &  15.164 (0.043) &     0.231 (0.043) &     0.071 (0.044) &  87 \\
  7:55:02.22 & $-$24:13:41.5 & 13.429 (0.026) &  0.412 (0.029) &  1.363  (0.057) &  13.414 (0.028) &     0.147 (0.028) &  $-$0.006 (0.034) &  111 \\
  7:55:02.48 & $-$24:16:44.6 & 14.570 (0.031) &  0.512 (0.039) &  1.146  (0.059) &  14.437 (0.030) &     0.321 (0.030) &     0.100 (0.036) &  203 \\
  7:55:03.00 & $-$24:18:04.9 & 14.901 (0.030) &  0.521 (0.039) &  1.412  (0.062) &  14.854 (0.035) &     0.240 (0.035) &     0.077 (0.041) &  245 \\
  7:55:03.21 & $-$24:14:11.7 & 13.706 (0.025) &  1.332 (0.033) &  2.833  (0.071) &  13.229 (0.028) &     1.088 (0.028) &     0.486 (0.033) &  120 \\
  7:55:03.98 & $-$24:10:38.6 & 14.719 (0.036) &  0.635 (0.046) &  1.300  (0.066) &  14.546 (0.038) &     0.403 (0.038) &     0.133 (0.046) &  10 \\
  7:55:04.29 & $-$24:18:19.0 & 15.363 (0.038) &  0.374 (0.050) &  1.396  (0.066) &  15.385 (0.039) &     0.115 (0.039) &     0.081 (0.043) &  255 \\
  7:55:04.47 & $-$24:16:51.0 & 10.756 (0.022) &  0.127 (0.022) &  1.122  (0.054) &  10.835 (0.023) &  $-$0.124 (0.023) &  $-$0.156 (0.029) &  206 \\

  7:55:05.18 & $-$24:15:15.7 & 12.968 (0.022) &  0.184 (0.025) &  0.954  (0.055) &  13.002 (0.023) &  $-$0.013 (0.023) &  $-$0.112 (0.029) &  154 \\
  7:55:05.28 & $-$24:17:06.4 & 14.072 (0.027) &  0.467 (0.031) &  1.147  (0.058) &  13.939 (0.028) &     0.311 (0.028) &     0.086 (0.034) &  212 \\
  7:55:05.34 & $-$24:16:39.2 & 12.037 (0.022) &  0.475 (0.023) &  1.174  (0.054) &  11.926 (0.024) &     0.285 (0.024) &     0.055 (0.032) &  200 \\
  7:55:05.48 & $-$24:16:34.9 & 15.288 (0.036) &  0.662 (0.051) &  1.489  (0.072) &  15.013 (0.034) &     0.519 (0.034) &     0.180 (0.043) &  196 \\
  7:55:05.56 & $-$24:15:07.0 & 15.041 (0.026) &  0.428 (0.038) &  1.437  (0.062) &  15.013 (0.033) &     0.162 (0.033) &     0.087 (0.047) &  149 \\
  7:55:06.22 & $-$24:17:54.2 & 14.965 (0.033) &  0.672 (0.042) &  1.320  (0.062) &  14.748 (0.039) &     0.486 (0.039) &     0.183 (0.047) &  238 \\
  7:55:06.34 & $-$24:11:21.6 & 13.898 (0.029) &  0.587 (0.034) &  1.414  (0.062) &  13.755 (0.036) &     0.369 (0.036) &     0.121 (0.043) &  37 \\
  7:55:06.99 & $-$24:18:58.5 & 14.668 (0.031) &  0.510 (0.040) &  1.212  (0.061) &  14.567 (0.038) &     0.340 (0.038) &     0.121 (0.038) &  280 \\
  7:55:07.12 & $-$24:18:27.3 & 14.529 (0.032) &  0.544 (0.042) &  1.190  (0.062) &  14.331 (0.032) &     0.410 (0.032) &     0.107 (0.041) &  261 \\
  7:55:07.57 & $-$24:18:42.2 & 12.963 (0.026) &  0.281 (0.027) &  1.192  (0.055) &  12.984 (0.032) &     0.057 (0.032) &  $-$0.020 (0.038) &  272 \\

  7:55:08.30 & $-$24:15:10.2 & 12.045 (0.023) &  0.152 (0.022) &  0.405  (0.054) &  12.095 (0.033) &  $-$0.060 (0.033) &  $-$0.078 (0.043) &  150 \\
  7:55:08.47 & $-$24:15:48.7 & 15.466 (0.037) &  0.538 (0.050) &  1.371  (0.073) &  15.341 (0.041) &     0.355 (0.041) &     0.159 (0.043) &  171 \\
  7:55:08.73 & $-$24:14:45.1 & 15.418 (0.031) &  0.510 (0.046) &  1.438  (0.073) &  15.269 (0.035) &     0.332 (0.035) &     0.103 (0.034) &  138 \\
  7:55:08.76 & $-$24:17:45.7 & 12.512 (0.025) &  0.357 (0.026) &  1.208  (0.054) &  12.489 (0.028) &     0.129 (0.028) &     0.003 (0.034) &  232 \\
  7:55:09.10 & $-$24:15:01.5 & 14.407 (0.094) &  1.253 (0.130) &  2.014  (0.144) &  14.040 (0.168) &     0.826 (0.168) &     0.361 (0.168) &  146 \\
  7:55:09.71 & $-$24:18:16.5 & 12.856 (0.027) &  0.223 (0.027) &  1.269  (0.055) &  12.884 (0.032) &  $-$0.003 (0.032) &  $-$0.042 (0.034) &  253 \\
  7:55:09.98 & $-$24:12:06.2 & 11.620 (0.028) &  1.688 (0.029) &  3.515  (0.062) &  10.976 (0.033) &     1.418 (0.033) &     0.627 (0.031) &  57 \\
  7:55:10.25 & $-$24:11:36.4 & 12.364 (0.031) &  0.162 (0.033) &  1.210  (0.057) &  12.427 (0.030) &  $-$0.078 (0.030) &  $-$0.099 (0.037) &  46 \\
  7:55:10.58 & $-$24:16:21.9 & 11.121 (0.021) &  0.122 (0.020) &  1.160  (0.053) &  11.216 (0.025) &  $-$0.145 (0.025) &  $-$0.139 (0.027) &  189 \\
  7:55:10.89 & $-$24:17:37.3 & 15.453 (0.040) &  0.456 (0.050) &  1.394  (0.071) &  15.405 (0.041) &     0.196 (0.041) &     0.115 (0.037) &  227 \\

  7:55:11.10 & $-$24:17:48.6 & 11.855 (0.025) &  0.126 (0.026) &  1.154  (0.054) &  11.953 (0.029) &  $-$0.135 (0.029) &  $-$0.115 (0.033) &  234 \\
  7:55:11.53 & $-$24:15:11.1 & 13.739 (0.022) &  0.437 (0.025) &  1.125  (0.055) &  13.672 (0.025) &     0.201 (0.025) &     0.032 (0.030) &  153 \\
  7:55:11.54 & $-$24:14:35.6 & 14.548 (0.030) &  0.568 (0.037) &  1.462  (0.061) &  14.394 (0.031) &     0.387 (0.031) &     0.230 (0.032) &  133 \\
  7:55:11.59 & $-$24:11:01.9 & 13.400 (0.032) &  0.069 (0.034) &  0.578  (0.059) &  13.440 (0.036) &  $-$0.096 (0.036) &  $-$0.121 (0.045) &  23 \\
  7:55:11.61 & $-$24:11:48.2 & 14.252 (0.030) &  0.634 (0.040) &  1.212  (0.060) &  14.083 (0.033) &     0.446 (0.033) &     0.133 (0.036) &  50 \\
  7:55:11.74 & $-$24:14:36.0 & 15.688 (0.038) &  0.588 (0.060) &  1.127  (0.066) &  15.527 (0.036) &     0.435 (0.036) &     0.116 (0.042) &  135 \\
  7:55:11.74 & $-$24:17:42.5 & 15.108 (0.036) &  0.585 (0.045) &  1.457  (0.066) &  14.907 (0.042) &     0.438 (0.042) &     0.211 (0.050) &  230 \\
  7:55:11.82 & $-$24:13:38.7 & 10.646 (0.022) &  0.122 (0.020) &  1.181  (0.053) &  10.739 (0.020) &  $-$0.141 (0.020) &  $-$0.111 (0.022) &  108 \\
  7:55:11.84 & $-$24:15:17.8 & 14.587 (0.028) &  0.600 (0.035) &  1.270  (0.061) &  14.428 (0.036) &     0.374 (0.036) &     0.136 (0.039) &  156 \\
  7:55:11.94 & $-$24:17:31.6 & 13.731 (0.029) &  0.513 (0.032) &  1.188  (0.056) &  13.610 (0.031) &     0.310 (0.031) &     0.150 (0.041) &  222 \\

  7:55:12.05 & $-$24:15:23.6 & 13.251 (0.022) &  0.423 (0.023) &  1.086  (0.055) &  13.152 (0.027) &     0.238 (0.027) &     0.071 (0.026) &  157 \\
  7:55:13.02 & $-$24:12:08.1 & 12.758 (0.026) &  1.089 (0.028) &  2.252  (0.059) &  12.407 (0.029) &     0.835 (0.029) &     0.343 (0.034) &  59 \\
  7:55:13.57 & $-$24:15:41.6 & 10.220 (0.022) &  1.102 (0.022) &  2.312  (0.054) &  9.972  (0.021) &     0.687 (0.025) &     0.299 (0.016) &  166 \\
  7:55:13.82 & $-$24:16:03.7 & 12.877 (0.025) &  0.267 (0.025) &  1.204  (0.055) &  12.891 (0.029) &     0.031 (0.029) &  $-$0.003 (0.034) &  179 \\
  7:55:13.92 & $-$24:14:35.4 & 15.431 (0.031) &  0.777 (0.048) &  1.412  (0.075) &  15.218 (0.038) &     0.559 (0.038) &     0.224 (0.043) &  134 \\
  7:55:14.15 & $-$24:18:47.9 & 15.354 (0.045) &  0.798 (0.063) &  1.548  (0.078) &  15.126 (0.058) &     0.602 (0.058) &     0.231 (0.063) &  275 \\
  7:55:15.36 & $-$24:12:28.8 & 14.699 (0.035) &  0.442 (0.042) &  1.384  (0.063) &  14.727 (0.030) &     0.118 (0.030) &     0.092 (0.042) &  73 \\
  7:55:15.80 & $-$24:13:19.1 & 11.262 (0.021) &  0.099 (0.021) &  1.106  (0.054) &  11.372 (0.021) &  $-$0.152 (0.021) &  $-$0.121 (0.031) &  94 \\
  7:55:15.87 & $-$24:17:09.9 & 14.329 (0.028) &  0.319 (0.033) &  1.244  (0.058) &  14.323 (0.033) &     0.133 (0.033) &     0.019 (0.038) &  214 \\
  7:55:15.90 & $-$24:12:11.0 & 13.515 (0.025) &  0.535 (0.030) &  1.161  (0.057) &  13.390 (0.028) &     0.332 (0.028) &     0.102 (0.033) &  62 \\

  7:55:16.15 & $-$24:16:34.1 & 11.410 (0.024) &  1.333 (0.027) &  2.973  (0.057) &  10.954 (0.027) &     1.054 (0.027) &     0.414 (0.027) &  197 \\
  7:55:16.77 & $-$24:12:42.1 & 15.294 (0.036) &  0.675 (0.048) &  1.376  (0.070) &  15.047 (0.043) &     0.590 (0.043) &     0.246 (0.048) &  78 \\
  7:55:16.90 & $-$24:12:12.9 & 12.525 (0.025) &  0.195 (0.025) &  1.209  (0.055) &  12.580 (0.030) &  $-$0.057 (0.030) &  $-$0.108 (0.040) &  66 \\
  7:55:17.30 & $-$24:12:09.5 & 12.752 (0.025) &  0.434 (0.030) &  1.331  (0.055) &  12.663 (0.030) &     0.238 (0.030) &     0.048 (0.031) &  61 \\
  7:55:17.52 & $-$24:11:22.1 & 15.211 (0.041) &  0.700 (0.050) &  1.206  (0.065) &  15.025 (0.039) &     0.485 (0.039) &     0.186 (0.051) &  38 \\
  7:55:17.93 & $-$24:16:18.0 & 12.670 (0.023) &  0.246 (0.024) &  1.108  (0.055) &  12.668 (0.029) &     0.056 (0.029) &  $-$0.024 (0.029) &  187 \\
  7:55:18.75 & $-$24:17:41.0 & 11.425 (0.026) &  0.087 (0.027) &  1.138  (0.058) &  11.519 (0.035) &  $-$0.143 (0.035) &  $-$0.161 (0.045) &  229 \\
  7:55:19.42 & $-$24:12:55.9 & 11.848 (0.024) &  1.167 (0.028) &  2.646  (0.056) &  11.419 (0.027) &     0.949 (0.027) &     0.353 (0.032) &  85 \\
  7:55:19.46 & $-$24:14:49.4 & 11.754 (0.021) &  0.242 (0.021) &  1.329  (0.054) &  11.748 (0.022) &     0.034 (0.022) &  $-$0.097 (0.021) &  143 \\
  7:55:19.46 & $-$24:17:56.6 & 11.881 (0.028) &  1.370 (0.031) &  2.720  (0.061) &  11.400 (0.040) &     1.110 (0.040) &     0.468 (0.048) &  242 \\

  7:55:20.12 & $-$24:10:55.7 & 11.592 (0.030) &  0.198 (0.035) &  1.267  (0.056) &  11.648 (0.034) &  $-$0.044 (0.034) &  $-$0.084 (0.044) &  19 \\
  7:55:20.59 & $-$24:13:28.1 & 13.391 (0.022) &  1.527 (0.029) &  3.067  (0.065) &  12.862 (0.027) &     1.238 (0.027) &     0.597 (0.028) &  103 \\
  7:55:20.63 & $-$24:18:36.0 & 15.323 (0.035) &  0.445 (0.046) &  1.363  (0.072) &  15.257 (0.051) &     0.238 (0.051) &     0.074 (0.064) &  269 \\
  7:55:20.73 & $-$24:14:44.1 & 12.682 (0.022) &  0.163 (0.022) &  1.218  (0.054) &  12.720 (0.025) &  $-$0.065 (0.025) &  $-$0.114 (0.023) &  139 \\
  7:55:20.78 & $-$24:17:52.1 & 13.501 (0.029) &  0.306 (0.032) &  1.280  (0.062) &  13.459 (0.042) &     0.120 (0.042) &     0.001 (0.049) &  239 \\
  7:55:20.92 & $-$24:17:55.1 & 11.926 (0.028) &  0.246 (0.030) &  1.179  (0.060) &  11.924 (0.040) &     0.052 (0.040) &  $-$0.043 (0.046) &  241 \\
  7:55:20.98 & $-$24:17:24.4 & 12.287 (0.024) &  0.149 (0.026) &  1.210  (0.058) &  12.354 (0.034) &  $-$0.098 (0.034) &  $-$0.141 (0.040) &  220 \\
  7:55:21.16 & $-$24:17:20.9 & 15.130 (0.035) &  0.649 (0.052) &  1.316  (0.066) &  14.926 (0.040) &     0.514 (0.040) &     0.233 (0.038) &  218 \\
  7:55:21.87 & $-$24:10:48.4 & 13.939 (0.031) &  1.138 (0.045) &  2.184  (0.067) &  13.510 (0.043) &     0.937 (0.043) &     0.407 (0.051) &  15 \\
  7:55:22.20 & $-$24:10:45.9 & 14.338 (0.036) &  0.552 (0.044) &  1.179  (0.060) &  14.189 (0.038) &     0.357 (0.038) &     0.095 (0.048) &  13 \\

  7:55:22.34 & $-$24:16:45.2 & 14.754 (0.030) &  0.722 (0.043) &  1.483  (0.063) &  14.544 (0.035) &     0.507 (0.035) &     0.125 (0.038) &  204 \\
  7:55:22.51 & $-$24:18:16.7 & 12.585 (0.030) &  0.480 (0.033) &  1.134  (0.062) &  12.463 (0.040) &     0.309 (0.040) &     0.072 (0.052) &  256 \\
  7:55:22.70 & $-$24:13:26.8 & 14.118 (0.026) &  0.293 (0.033) &  1.023  (0.057) &  14.086 (0.028) &     0.113 (0.028) &  $-$0.004 (0.032) &  102 \\
  7:55:23.06 & $-$24:14:22.1 & 14.967 (0.032) &  0.595 (0.041) &  1.177  (0.062) &  14.843 (0.033) &     0.401 (0.033) &     0.090 (0.036) &  129 \\
  7:55:23.09 & $-$24:10:58.4 & 13.123 (0.030) &  0.184 (0.036) &  0.691  (0.056) &  13.157 (0.037) &  $-$0.044 (0.037) &  $-$0.072 (0.046) &  21 \\
  7:55:23.11 & $-$24:15:45.8 & 13.526 (0.024) &  0.353 (0.026) &  1.159  (0.057) &  13.471 (0.030) &     0.169 (0.030) &  $-$0.016 (0.031) &  170 \\
  7:55:23.43 & $-$24:18:29.0 & 13.412 (0.031) &  0.327 (0.037) &  1.198  (0.063) &  13.359 (0.042) &     0.131 (0.042) &  $-$0.003 (0.051) &  265 \\
  7:55:24.00 & $-$24:17:51.0 & 15.392 (0.036) &  0.458 (0.048) &  1.457  (0.068) &  15.278 (0.044) &     0.280 (0.044) &     0.065 (0.051) &  236 \\
  7:55:24.18 & $-$24:14:51.3 & 12.229 (0.022) &  0.212 (0.024) &  1.210  (0.055) &  12.224 (0.027) &     0.032 (0.027) &  $-$0.072 (0.029) &  144 \\
  7:55:24.20 & $-$24:18:40.2 & 14.951 (0.042) &  0.682 (0.054) &  1.280  (0.064) &  14.726 (0.051) &     0.534 (0.051) &     0.195 (0.050) &  273 \\

  7:55:24.20 & $-$24:18:51.8 & 14.548 (0.039) &  0.571 (0.047) &  1.374  (0.069) &  14.352 (0.045) &     0.430 (0.045) &     0.169 (0.051) &  278 \\
  7:55:24.36 & $-$24:17:33.3 & 15.027 (0.031) &  0.625 (0.039) &  1.203  (0.069) &  14.900 (0.044) &     0.363 (0.044) &     0.155 (0.051) &  224 \\
  7:55:24.48 & $-$24:18:00.1 & 12.755 (0.047) &  0.582 (0.061) &  1.245  (0.080) &  12.618 (0.096) &     0.386 (0.096) &     0.010 (0.101) &  244 \\
  7:55:24.91 & $-$24:16:50.7 & 11.425 (0.025) &  0.110 (0.026) &  1.180  (0.058) &  11.462 (0.032) &  $-$0.084 (0.032) &  $-$0.133 (0.038) &  208 \\
  7:55:25.25 & $-$24:10:34.3 & 14.650 (0.032) &  0.612 (0.045) &  1.273  (0.061) &  14.472 (0.039) &     0.418 (0.039) &     0.189 (0.049) &  9 \\
  7:55:25.54 & $-$24:10:34.1 & 14.772 (0.040) &  0.529 (0.048) &  1.174  (0.064) &  14.604 (0.047) &     0.335 (0.047) &     0.131 (0.052) &  8 \\
  7:55:25.73 & $-$24:15:42.6 & 15.285 (0.030) &  0.781 (0.048) &  1.432  (0.066) &  15.103 (0.040) &     0.531 (0.040) &     0.191 (0.044) &  167 \\
  7:55:26.16 & $-$24:12:42.7 & 14.575 (0.031) &  1.242 (0.051) &  2.242  (0.073) &  14.127 (0.037) &     1.001 (0.037) &     0.416 (0.038) &  79 \\
  7:55:26.81 & $-$24:18:57.2 & 11.435 (0.037) &  0.102 (0.043) &  0.642  (0.065) &  11.439 (0.047) &  $-$0.038 (0.047) &  $-$0.140 (0.060) &  283 \\
  7:55:26.86 & $-$24:12:15.2 & 15.404 (0.034) &  0.691 (0.053) &  1.399  (0.068) &  15.167 (0.046) &     0.475 (0.046) &     0.147 (0.051) &  68 \\

  7:55:27.59 & $-$24:13:30.5 & 14.556 (0.030) &  0.475 (0.042) &  1.068  (0.060) &  14.403 (0.038) &     0.299 (0.038) &     0.101 (0.041) &  104 \\
  7:55:27.61 & $-$24:14:44.4 & 13.044 (0.024) &  0.267 (0.028) &  1.237  (0.056) &  13.003 (0.032) &     0.088 (0.032) &  $-$0.083 (0.033) &  140 \\
  7:55:27.82 & $-$24:11:55.0 & 12.890 (0.027) &  0.331 (0.037) &  1.138  (0.058) &  12.807 (0.040) &     0.169 (0.040) &     0.023 (0.041) &  53 \\
  7:55:28.14 & $-$24:12:36.6 & 15.513 (0.036) &  0.605 (0.052) &  1.364  (0.074) &  15.287 (0.050) &     0.401 (0.050) &     0.154 (0.052) &  77 \\
  7:55:28.30 & $-$24:16:15.0 & 13.285 (0.025) &  0.440 (0.028) &  0.903  (0.057) &  13.137 (0.030) &     0.291 (0.030) &     0.072 (0.037) &  185 \\
  7:55:29.15 & $-$24:14:57.4 & 13.493 (0.026) &  0.474 (0.034) &  1.187  (0.058) &  13.344 (0.039) &     0.323 (0.039) &     0.030 (0.042) &  145 \\
  7:55:29.39 & $-$24:14:22.6 & 13.008 (0.023) &  0.265 (0.027) &  1.194  (0.057) &  12.964 (0.032) &     0.093 (0.032) &  $-$0.104 (0.034) &  130 \\
  7:55:29.66 & $-$24:17:01.8 & 13.312 (0.030) &  0.408 (0.033) &  1.180  (0.059) &  13.194 (0.035) &     0.249 (0.035) &     0.024 (0.045) &  211 \\
  7:55:29.78 & $-$24:12:01.1 & 13.115 (0.024) &  0.280 (0.037) &  1.049  (0.056) &  13.054 (0.041) &     0.123 (0.041) &  $-$0.006 (0.040) &  56 \\
  7:55:31.14 & $-$24:10:29.9 & 14.105 (0.030) &  0.694 (0.044) &  1.351  (0.060) &  13.874 (0.048) &     0.471 (0.048) &     0.161 (0.054) &  5 \\

  7:55:32.13 & $-$24:14:44.8 & 12.990 (0.024) &  0.282 (0.028) &  1.249  (0.056) &  12.960 (0.033) &     0.084 (0.033) &  $-$0.072 (0.035) &  141 \\
  7:55:32.63 & $-$24:14:08.3 & 14.762 (0.028) &  0.627 (0.049) &  1.123  (0.063) &  14.604 (0.041) &     0.419 (0.041) &     0.122 (0.042) &  121 \\
  7:55:33.46 & $-$24:18:06.4 & 15.454 (0.050) &  0.554 (0.064) &  1.176  (0.075) &  15.274 (0.054) &     0.371 (0.054) &     0.124 (0.065) &  249 \\
  7:55:33.56 & $-$24:18:21.7 & 12.936 (0.038) &  1.683 (0.043) &  3.603  (0.080) &  12.261 (0.044) &     1.440 (0.044) &     0.666 (0.056) &  260 \\
\enddata
\tablenotetext{a}{Values in parentheses are 1-$\sigma$ random errors, including
uncertainty of nightly fit to standards.}
\tablenotetext{b}{Equinox J2000.}
\end{deluxetable}

\begin{deluxetable}{crc}
\tablewidth{0pc}
\tablecaption{NGC 2482 Cross-Identifications\label{ids}}
\tablehead{   \colhead{ID$_M$\tablenotemark{a}} & 
\colhead{ID$_K$\tablenotemark{b}} & \colhead{Luminosity Class\tablenotemark{c}} } 
\startdata
16 & 256 & V \\
17 & 241/239 & V \\
18 & 242 & n \\
19 & 229 & V \\
20 & 220/218 & V \\
21 & 197 & n \\
22 & 187 & V \\
23 & 166 & III \\
24 & 189 & V \\
25 & 200 & V \\
26 & 206 & III \\
27 & 177 & V \\
29 & 112 & V \\
30 & 108 & III \\
32 &  57 & n \\
33 &  94 & V \\
34 &  85 & n \\
35 & 143 & n \\
36 & 208 & V \\
\enddata
\tablenotetext{a}{ID in \citet{MofVog75}.}
\tablenotetext{b}{ID in Table \ref{photometry} of this paper.}
\tablenotetext{c}{V = main sequence; III = giant; n = non-cluster member.
From \citet{MofVog75}.  Two of their stars (17 and 20) actually are comprised
of two components. Their star number 35 is erroneously designated a non-member.}
\end{deluxetable}


\begin{deluxetable}{lcccccr}
\tablecolumns{7} 
\tablewidth{0pc} 
\tabletypesize{\scriptsize}
\rotate \tablecaption{Derived Results\tablenotemark{a} \label{results}} 
\tablehead{
\colhead{Cluster} & \colhead{log $t$ (yrs)} & \colhead{D(pc)} &
\colhead{E($B-V$)} & \colhead{R$_V$\tablenotemark{b}} & \colhead{Z} &
\colhead{[Fe/H]\tablenotemark{c}} } 
\startdata
AH03 J0748$-$26.9   & 7.80 (0.24) & 4440 (353) & 0.390 (0.025) & 2.715 (0.263) & 0.015 (0.006) &  $-$0.10 (+0.15, $-$0.22) \\
NGC 2482            & 8.65 (0.09) & 1194 (112) & 0.094 (0.040) & \ldots        & 0.015 (0.005) &  $-$0.10 (+0.13, $-$0.18) \\
NGC 2482\tablenotemark{d} & 8.57 (0.09) & 1349 (117) & 0.086 (0.029) & \ldots  & 0.022 (0.006) &    +0.06 (+0.10, $-$0.14) \\
Ruprecht 42         & 8.60 (0.13) & 5457 (418) & 0.388 (0.061) & 4.059 (0.875) & 0.009 (0.003) &  $-$0.33 (+0.13, $-$0.17) \\
Ruprecht 153        & 8.65 (0.17) & 1848 (157) & 0.206 (0.042) & 4.111 (0.605) & 0.024 (0.006) &    +0.10 (+0.10, $-$0.13) \\
Ruprecht 154        & 9.35 (0.28) & 1783 (175) & 0.087 (0.024) & \ldots        & 0.027 (0.004) &    +0.15 (+0.06, $-$0.07) \\  
Ruprecht 51         & 9.00 (0.07) & 2514 (245) & 0.189 (0.041) & 3.415 (0.684) & 0.021 (0.005) & +0.04 (+0.09, $-$0.12) \\ 
\enddata 
\tablenotetext{a}{Values in parentheses are 1-$\sigma$ error bars. Except where noted, the results are derived 
from PSF photometry.}
\tablenotetext{b}{If the color excess is less than 0.1 mag, we do not derive R$_V$.} 
\tablenotetext{c}{[Fe/H] = log(Z/Z$_{\odot}$). The solar metallicity is taken to be 0.019.} 
\tablenotetext{d}{Results from aperture photometry.} \end{deluxetable}

\begin{deluxetable}{lcrcrcrcrc}
\tablecolumns{10}
\tablewidth{0pc}
\tablecaption{Exposure Times and Seeing\tablenotemark{a}\label{exptimes}}
\tablehead{ \colhead{Cluster} & \colhead{UT Date} &
\colhead{$t_u$} & \colhead{$s_U$} &  
\colhead{$t_B$} & \colhead{$s_B$} &  
\colhead{$t_V$} & \colhead{$s_V$} &  
\colhead{$t_g$} & \colhead{$s_g$} }
\startdata
NGC 2482          & 010612 &  24 & 2.7 &   3  &  2.6 &   3 & 2.4 &  3 & 2.2 \\
                  & 122112 &     &     & 150  &  3.3 & 120 & 3.4 &    &     \\ 
\hline
Ruprecht 51       & 123113 & 120 & 2.8 & 120  &  2.6 &  90 & 2.6 & 40 & 2.5 \\
                  &        &     &     &  20  &  2.4 &   6 & 2.4 &  6 & 2.5 \\
\hline
AH03 J0748$-$26.9 & 010114 & 100 & 2.9 &  70  &  3.2 &  50 & 2.8 &    &     \\
                  &        &  30 & 3.0 &   7  &  2.5 &   7 & 2.7 &  7 & 3.2 \\
\hline
Ruprecht 42       & 010114 & 100 & 2.7 &  50  &  3.1 &  40 & 2.8 & 30 & 2.4 \\
                  &        &     &     &   5  &  2.9 &   5 & 2.4 &  5 & 2.4 \\
\hline
Ruprecht 153      & 010314 & 120 & 3.1 &  60  &  3.2 &  40 & 3.1 & 25 & 3.0 \\
                  &        &  30 & 3.1 &   5  &  3.2 &   5 & 2.7 &  5 & 2.8 \\
\hline
Ruprecht 154      & 010414 & 120 & 3.0 &  60 &   3.2 &  60 & 3.0 & 25 & 2.9 \\
                  &        &  30 & 3.2 &   5 &   3.1 &   5 & 2.9 &  5 & 2.7 \\   
\enddata
\tablenotetext{a}{The UT date is in MMDDYY format.
The nominal exposure times are given in seconds.  The
seeing (FWHM) is measured in pixels, where 1 px = 0.435 arc seconds.
Nominal 3 second $r$ and $i$ exposures were also taken on 6 January 2012.  The
seeing was 2.2 and 1.9 px, respectively, for those exposures.}
\end{deluxetable}

\clearpage

\begin{deluxetable}{cccrrc}
\tablecolumns{6}
\tablewidth{0pc}
\tablecaption{Photometric Transformation Parameters from Standard Stars\tablenotemark{a}\label{transform_params}}
\tablehead{ \colhead{UT Date} &
\colhead{Index} & \colhead{Extinction Term} &  
\colhead{Color Term} & \colhead{Zero Point} &  
\colhead{RMS} }
\startdata
010612 & $V$                     &  0.12            & $-$0.052 (0.015) & $-$2.903 (0.012) & $\pm$ 0.010 \\
       & $B$                     &  0.25            &   +0.042 (0.013) & $-$2.867 (0.010) & $\pm$ 0.017 \\
       & $u^{\prime}-g^{\prime}$ &  0.43            &    1.004 (0.020) & $-$1.605 (0.073) & $\pm$ 0.052 \\
       & $r^{\prime}$            &  0.09            &   +0.008 (0.014) & $-$2.471 (0.007) & $\pm$ 0.015 \\
       & $g^{\prime}-r^{\prime}$ &  0.04            &    0.961 (0.014) &    0.237 (0.008) & $\pm$ 0.016 \\
       & $r^{\prime}-i^{\prime}$ &  0.025           &    0.945 (0.019) &    0.383 (0.005) & $\pm$ 0.010 \\
\hline
122112 & $V$                     &  0.105 (0.026)   & $-$0.059 (0.016) & $-$3.086 (0.033) & $\pm$ 0.022 \\
       & $B$                     &  0.197 (0.030)   &   +0.063 (0.018) & $-$3.106 (0.038) & $\pm$ 0.024 \\
\hline
123113 & $V$                     &  0.103 (0.015)   & $-$0.089 (0.011) & $-$1.310 (0.021) & $\pm$ 0.019 \\
       & $B$                     &  0.238 (0.015)   &   +0.077 (0.011) & $-$1.289 (0.021) & $\pm$ 0.019 \\
       & $u^{\prime}-g^{\prime}$ &  0.271 (0.059)   &    1.000 (0.018) & $-$1.956 (0.096) & $\pm$ 0.041 \\
\hline
010114 & $V$                     &  0.146 (0.018)   & $-$0.091 (0.009) & $-$1.253 (0.025) & $\pm$ 0.015 \\
       & $B$                     &  0.259 (0.019)   &   +0.083 (0.009) & $-$1.263 (0.026) & $\pm$ 0.015 \\
       & $u^{\prime}-g^{\prime}$ &  0.350 (0.057)   &    1.003 (0.015) & $-$1.868 (0.107) & $\pm$ 0.037 \\
\hline
010314 & $V$                     &  0.14            & $-$0.104 (0.012) & $-$1.257 (0.007) & $\pm$ 0.017 \\
       & $B$                     &  0.24            &   +0.064 (0.013) & $-$1.289 (0.008) & $\pm$ 0.020 \\
       & $u^{\prime}-g^{\prime}$ &  0.311           &    1.002 (0.021) & $-$1.916 (0.077) & $\pm$ 0.051 \\
\hline
010414 & $V$                     &  0.108           & $-$0.095 (0.008) & $-$1.340 (0.008) & $\pm$ 0.020 \\
       & $B$                     &  0.218           &   +0.073 (0.018) & $-$1.365 (0.010) & $\pm$ 0.023 \\
       & $u^{\prime}-g^{\prime}$ &  0.311           &    1.035 (0.014) & $-$2.032 (0.047) & $\pm$ 0.026 \\
\enddata
\tablenotetext{a}{UT date is in MMDDYY format.  Extinction terms are measured in magnitudes
per airmass.  If no error bars are given, these are assumed values; on these nights all
objects were observed over a small range of airmass.  The $B$, $V$, and $r^{\prime}$ color
terms scale instrumental $b-v$, $b-v$, and $g-r$ instrumental colors, respectively.
Only aperture photometry was done with imagery of 21 December 2012.  All other zeropoints 
refer to PSF photometry.}
\end{deluxetable}

\clearpage


\figcaption[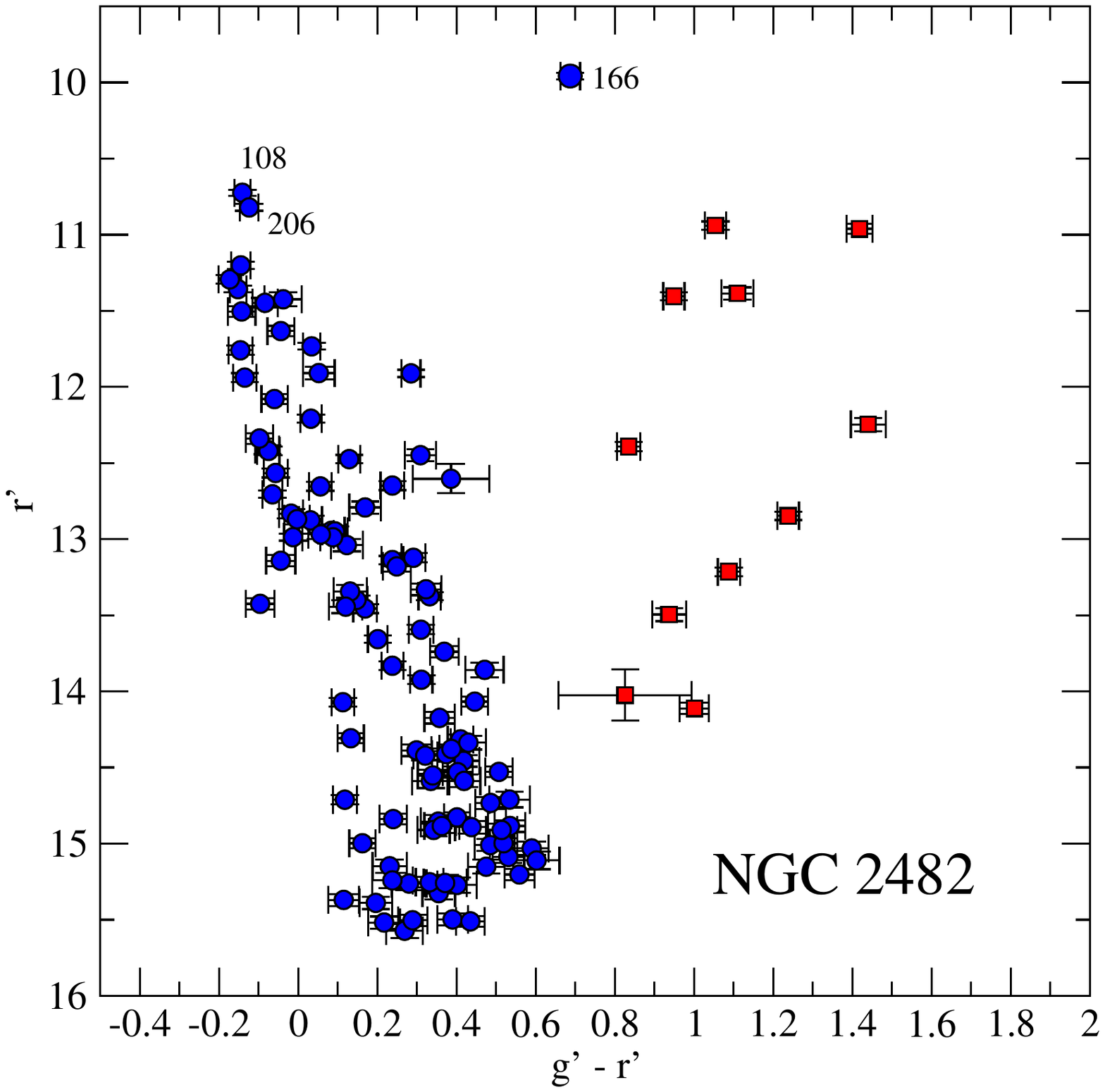]{$r^{\prime}$ vs. $g^{\prime} - r^{\prime}$
color-magnitude diagram of 114 stars in the field of NGC 2482.  
Obvious non-members are plotted as red squares. \label{ngc2482_rgr}
}

\figcaption[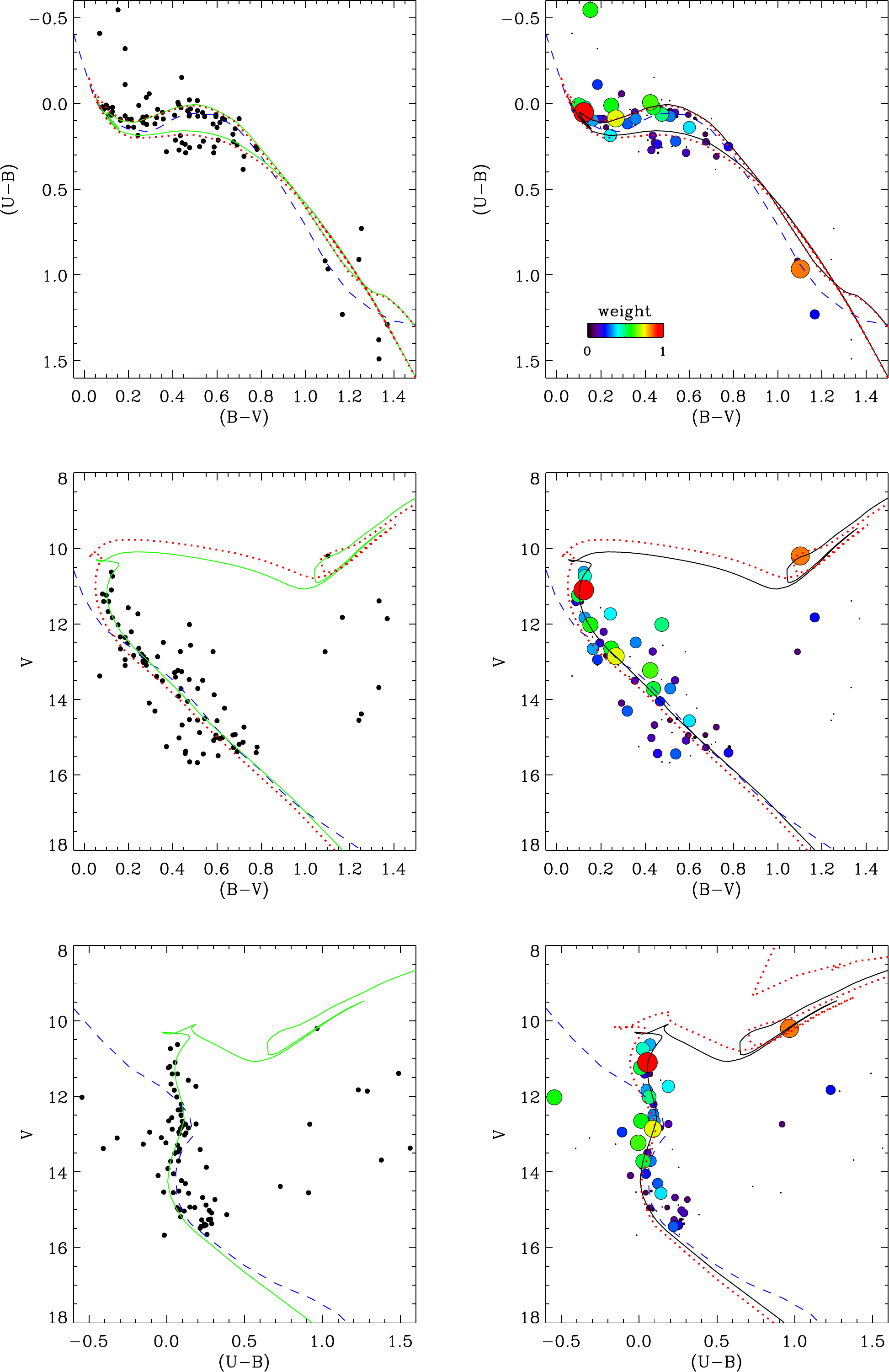]{Color-color and color-magnitude
magnitude diagrams of NGC 2482.  In each panel the zero age main sequence is
represented by a dashed line.  The solid line is the best fit isochrone. 
The red dotted lines correspond to isochrones based on the parameters
given by \citet{Kha_etal13}.
The likelihood of a star being a cluster member is color coded
according to the key in the top right-hand panel, and is also proportional
to the size of the points.\label{ngc2482}
}

\figcaption[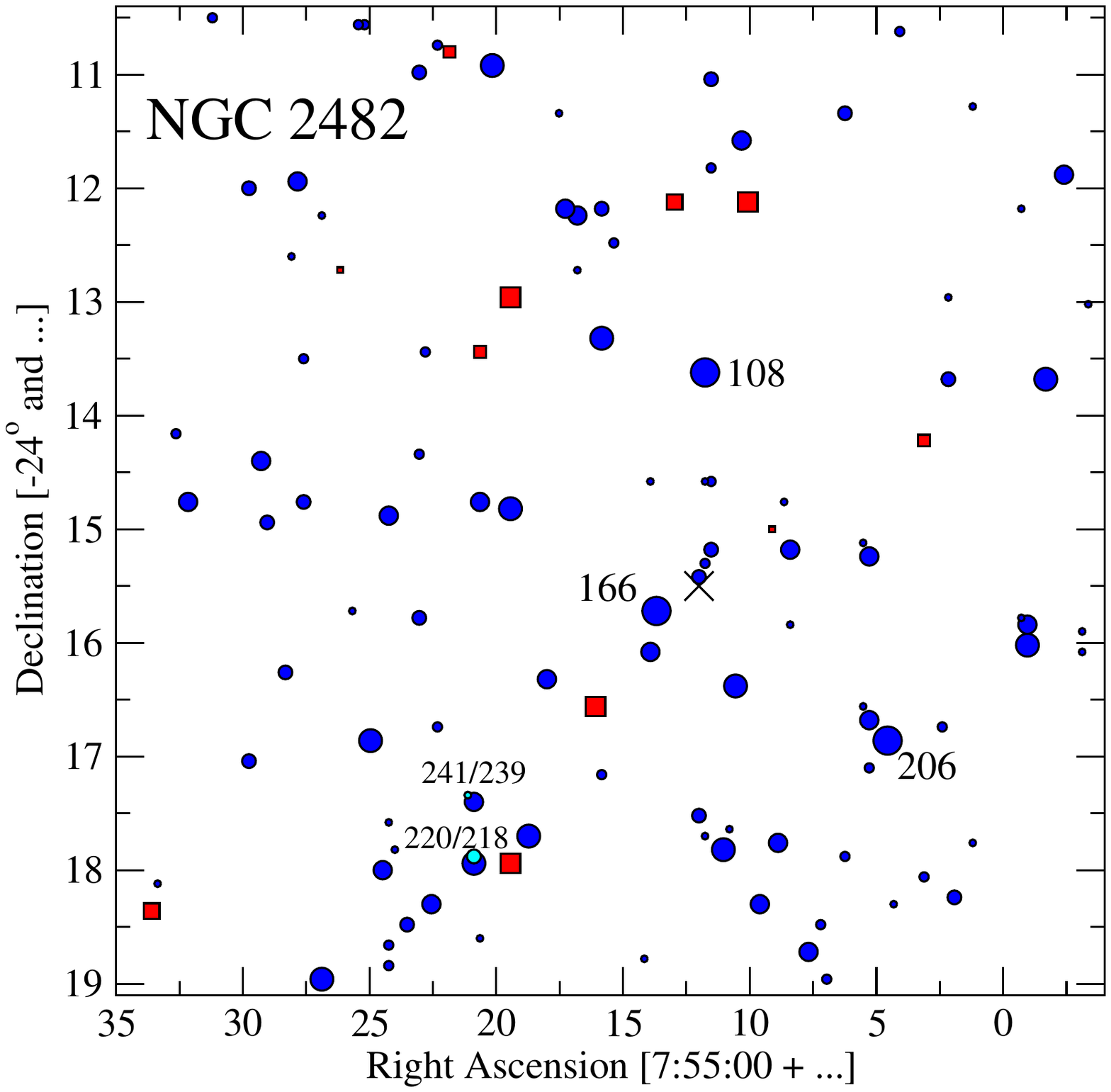]{A finder for the NGC 2482 stars
  discussed in this paper. North is up, east to the left.  The X marks
  the nominal location of the cluster center from
  \citet{Dia_etal02}. We sorted the stars according to the $V$-band
photometry, and binned the data from 10.0 to 10.99, 11.00 to 11.99 etc.
The sizes of the points in this figure decrease with the brightness
of the stars.  As in Fig. \ref{ngc2482_rgr}, likely cluster 
  members are plotted as round dots; likely giant stars at some 
  other distance are plotted as red squares.
  Stars 17 and 20 of \citet {MofVog75} have close companions
  visible in the CCD images; these companions are plotted as
  cyan-colored dots.  Star 166 is a bona fide giant in the cluster.
  Stars 108 and 206 are also past the turnoff point.
\label{finder}
}

\figcaption[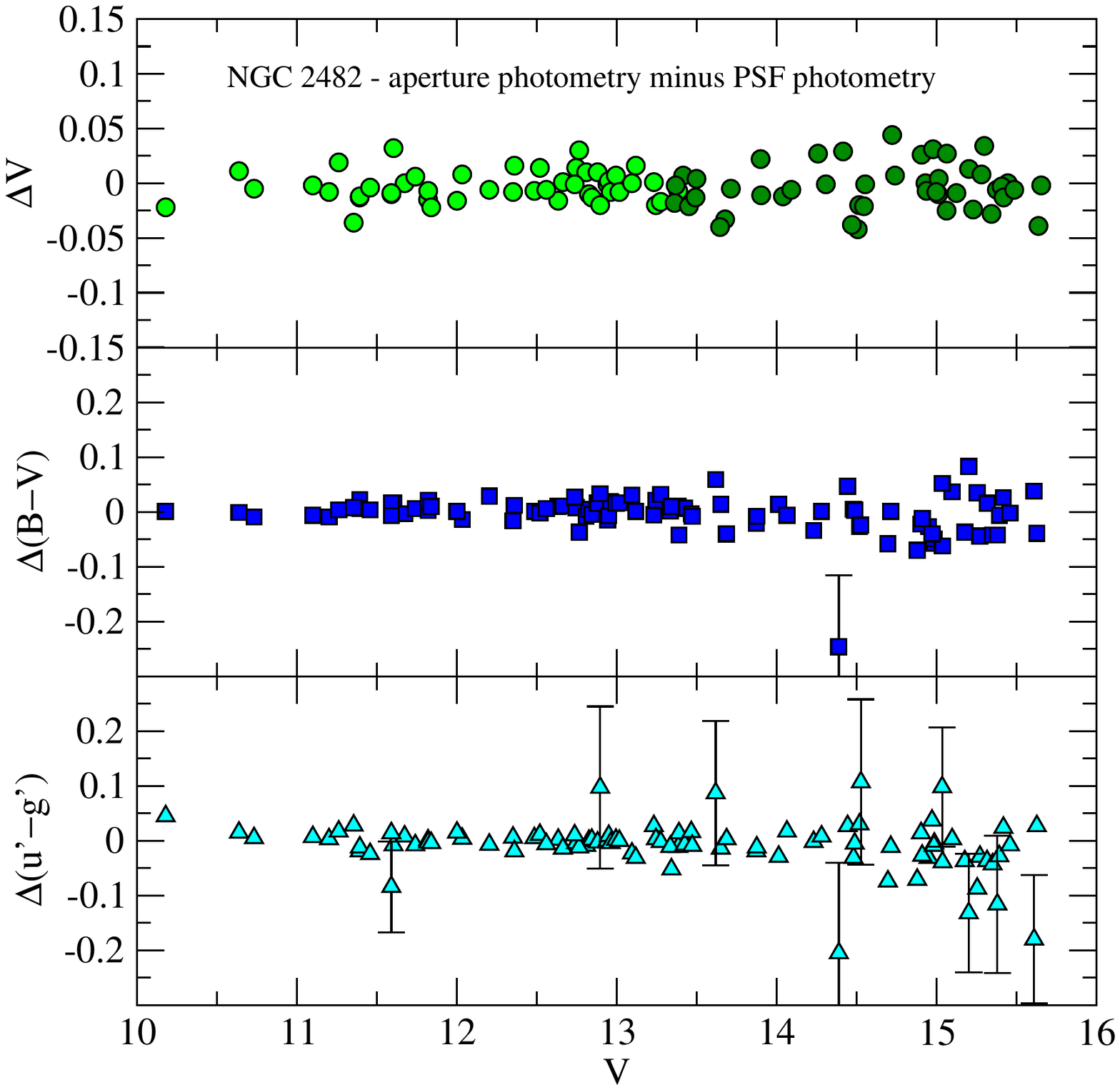]{Comparison of photometry of NGC 2482
done with aperture photometry and with PSF photometry. 
For the top two panels we used $BV$ aperture photometry of 6 January
2012 (short exposures) for stars brighter than $V \approx$ 13.2
and aperture photometry of 21 December 2012 (long exposures)
for stars fainter than $V \approx$ 13.2.  Sloan $u$-band
and $g$-band images were only taken on 6 January 2012.  This
figure illustrates the consistency of the photometry taken
with short and long exposure times on two different nights, and 
the consistency of photometry derived using instrumental
magnitudes obtained with two different methods.
\label{pm_comp}
}

\figcaption[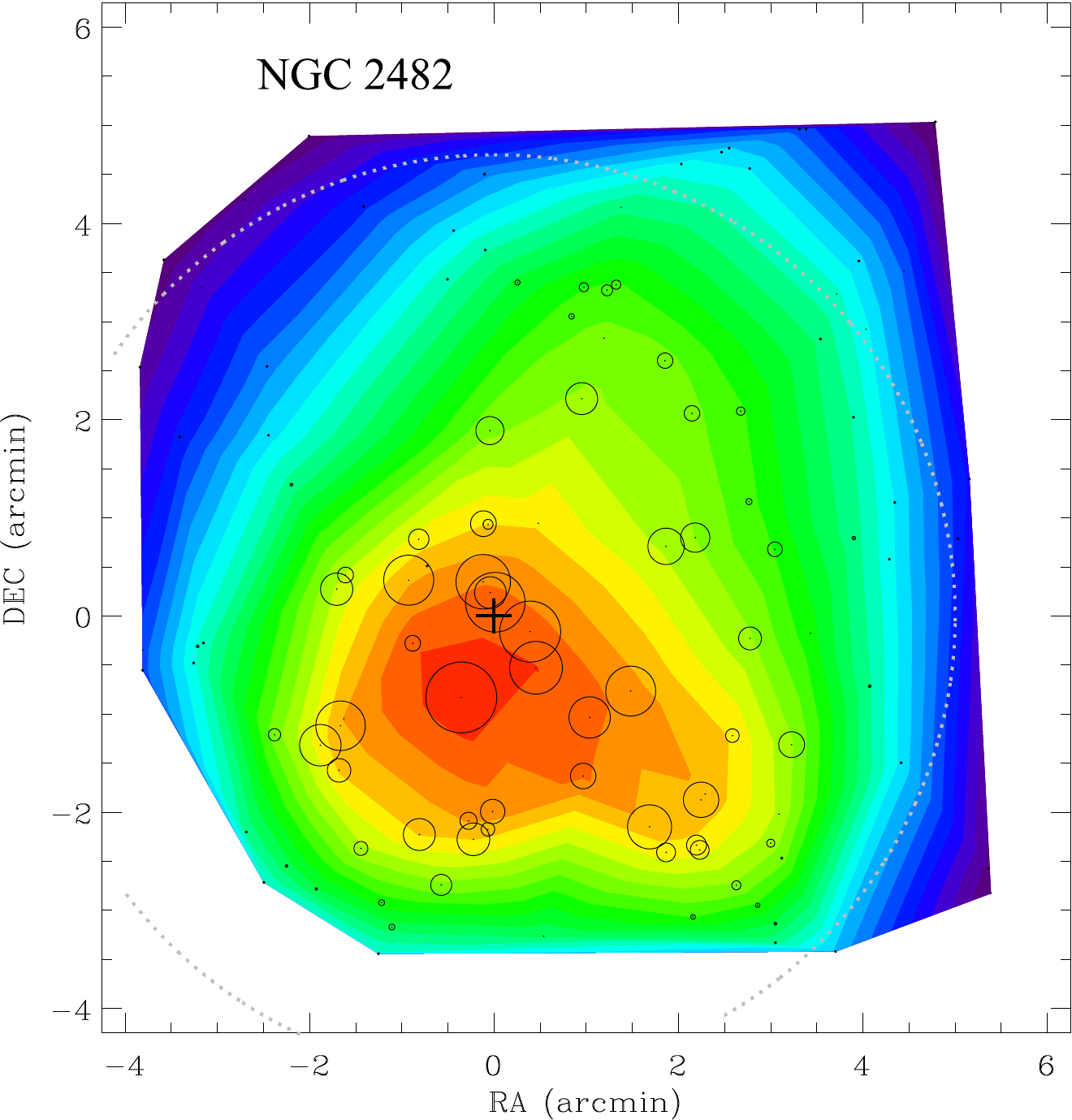]{Comparison of the density maps of
our six clusters.  For Ruprecht 153 the map was
obtained with a Gaussian kernel of $\sigma$ = 2.5\arcmin.
For the other clusters we used a kernal of 1\arcmin.\label{density}
}

\figcaption[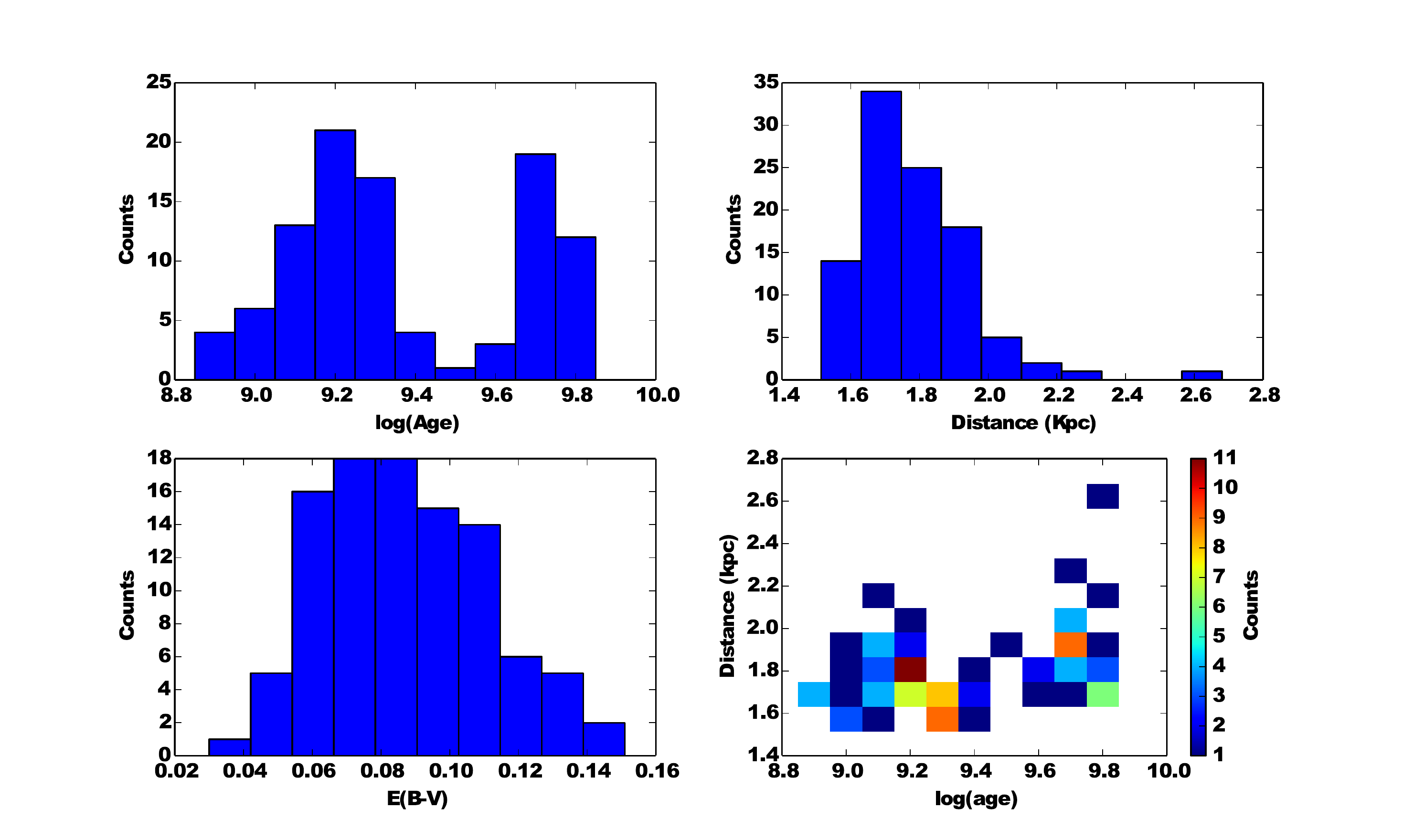]{Histograms of relative likelihood
of stellar age, distance, and color excess for the presumed
cluster Ruprecht 154.  The upper left and lower
right panels suggest that there are two populations of stars
along this line of sight. \label{rup154-hists}
}

\figcaption[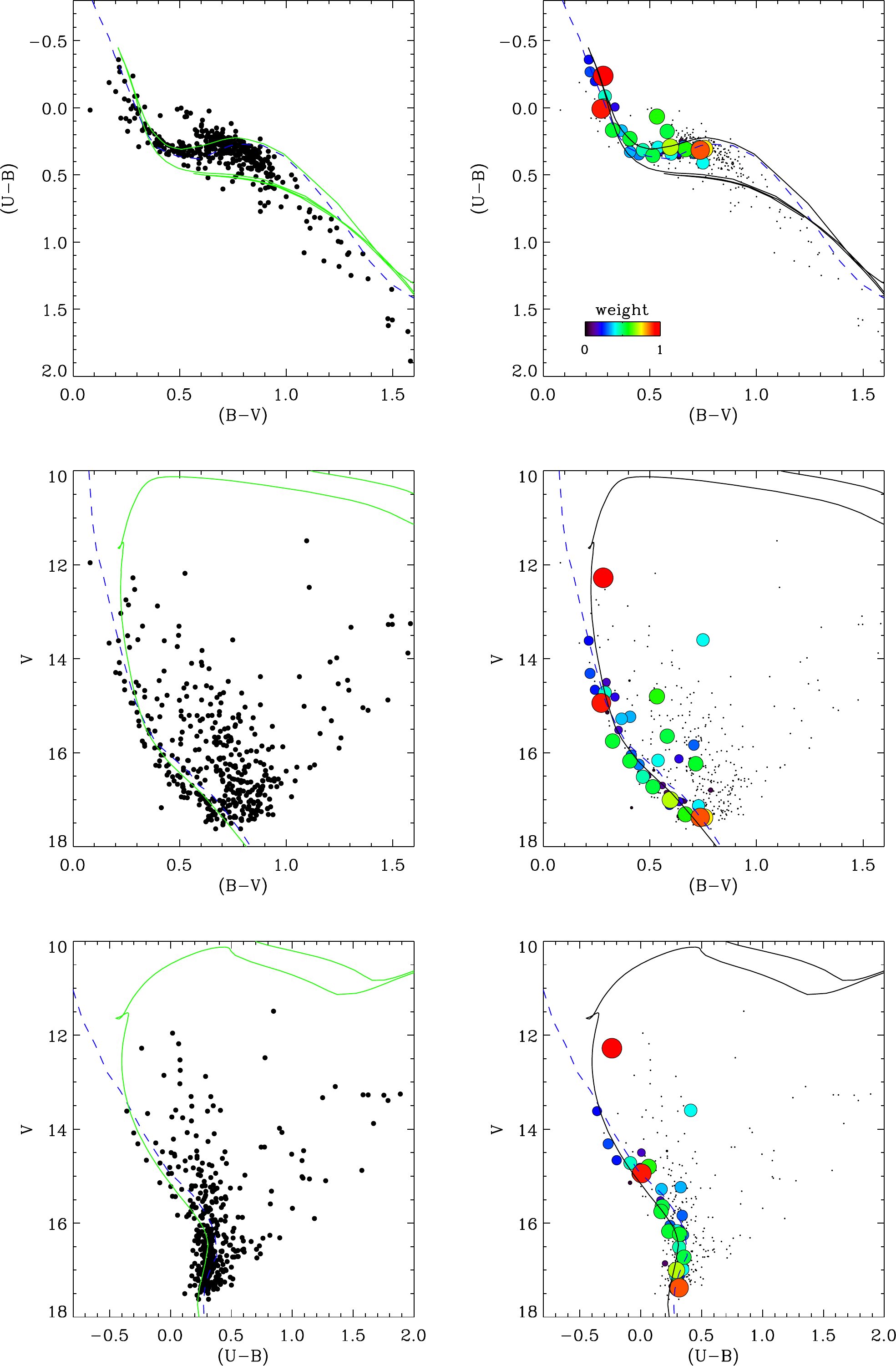]{Same as Fig. \ref{ngc2482}, but
for AH03 J0748$-$26.9.\label{ah03}
}

\figcaption[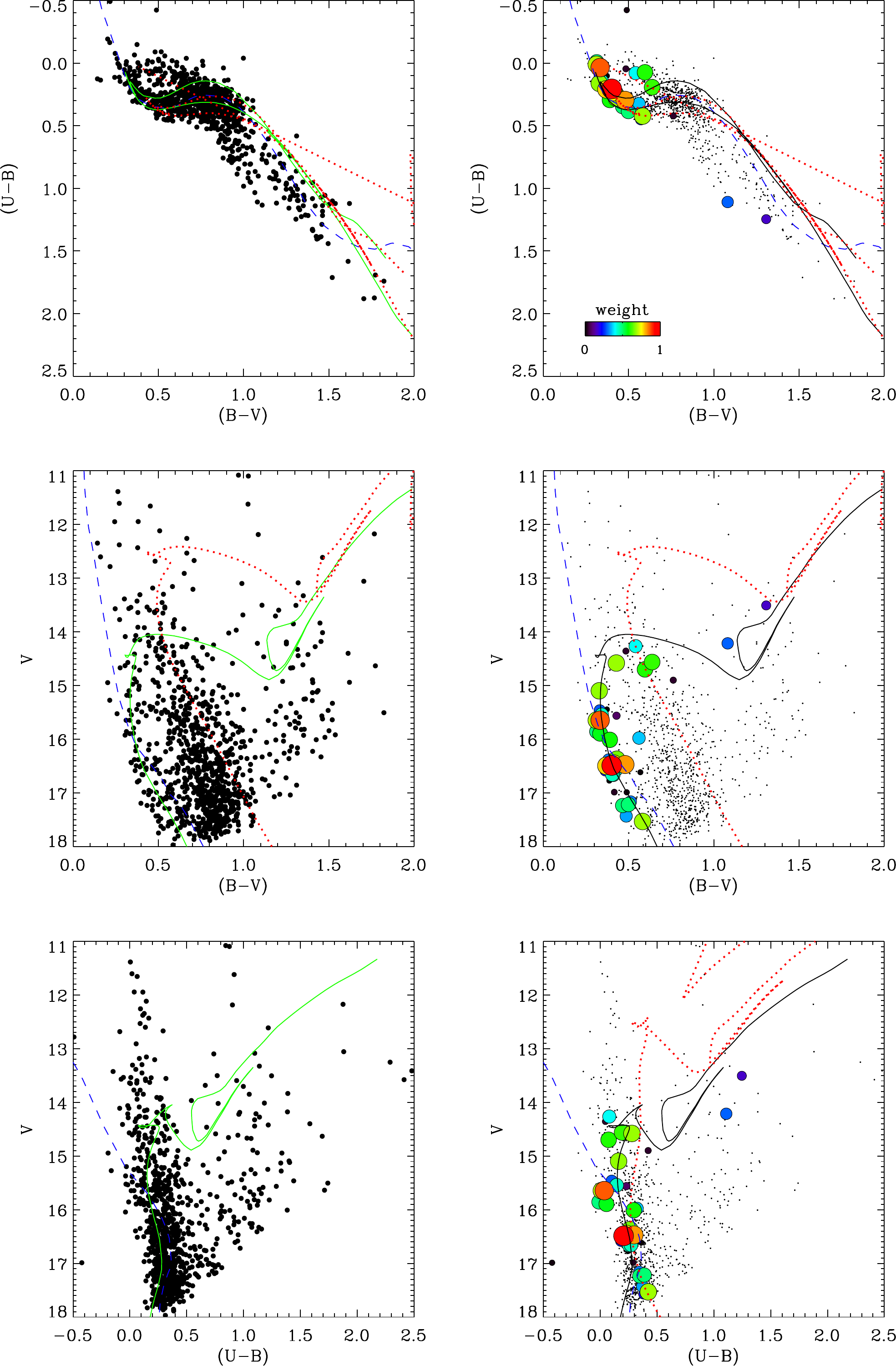]{Same as Fig. \ref{ngc2482}, but
for Ruprecht 42.\label{rup42}
}

\figcaption[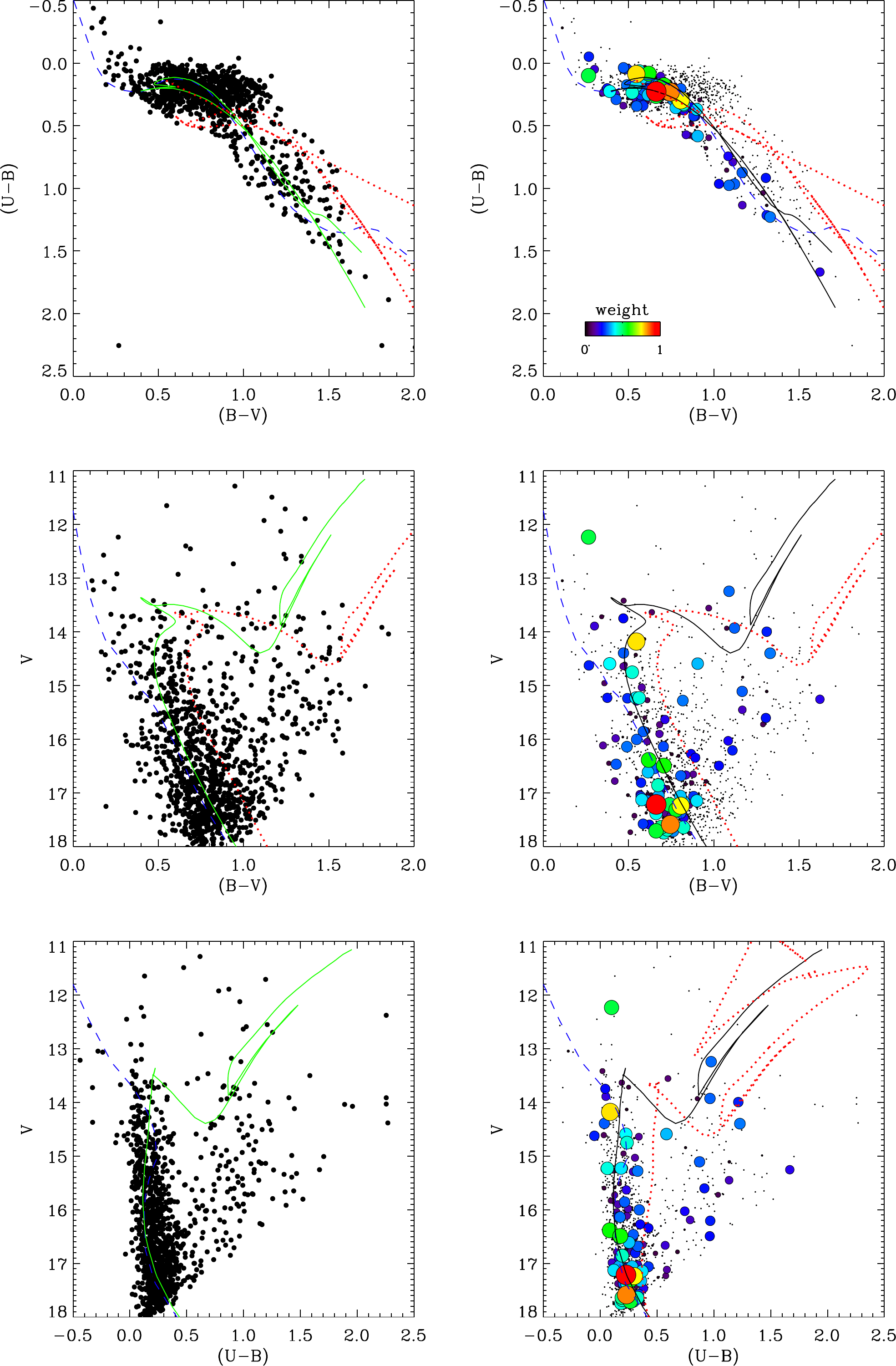]{Same as Fig. \ref{ngc2482}, but
for Ruprecht 51.\label{rup51}
}

\figcaption[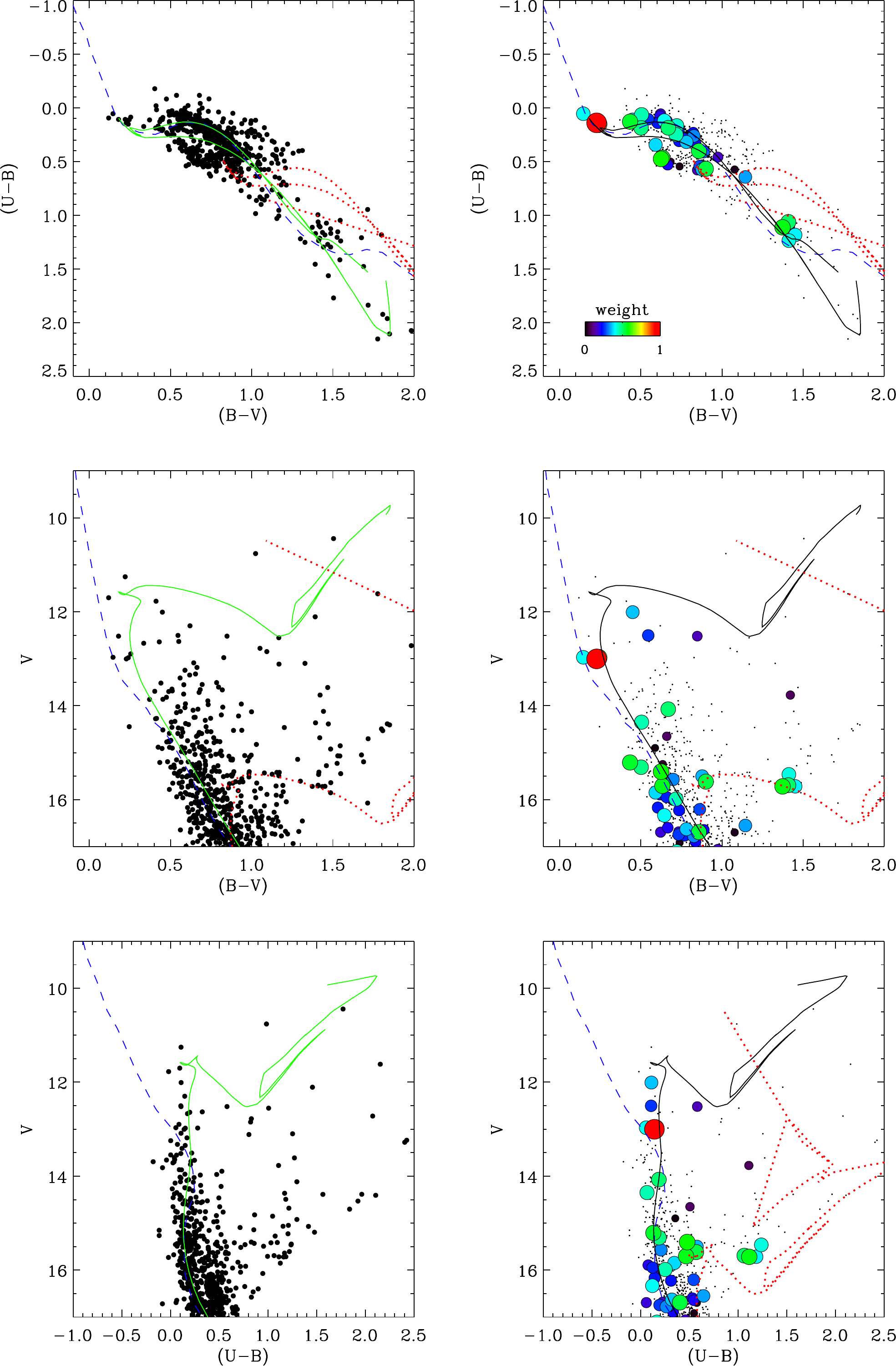]{Same as Fig. \ref{ngc2482}, but
for Ruprecht 153.\label{rup153}
}

\figcaption[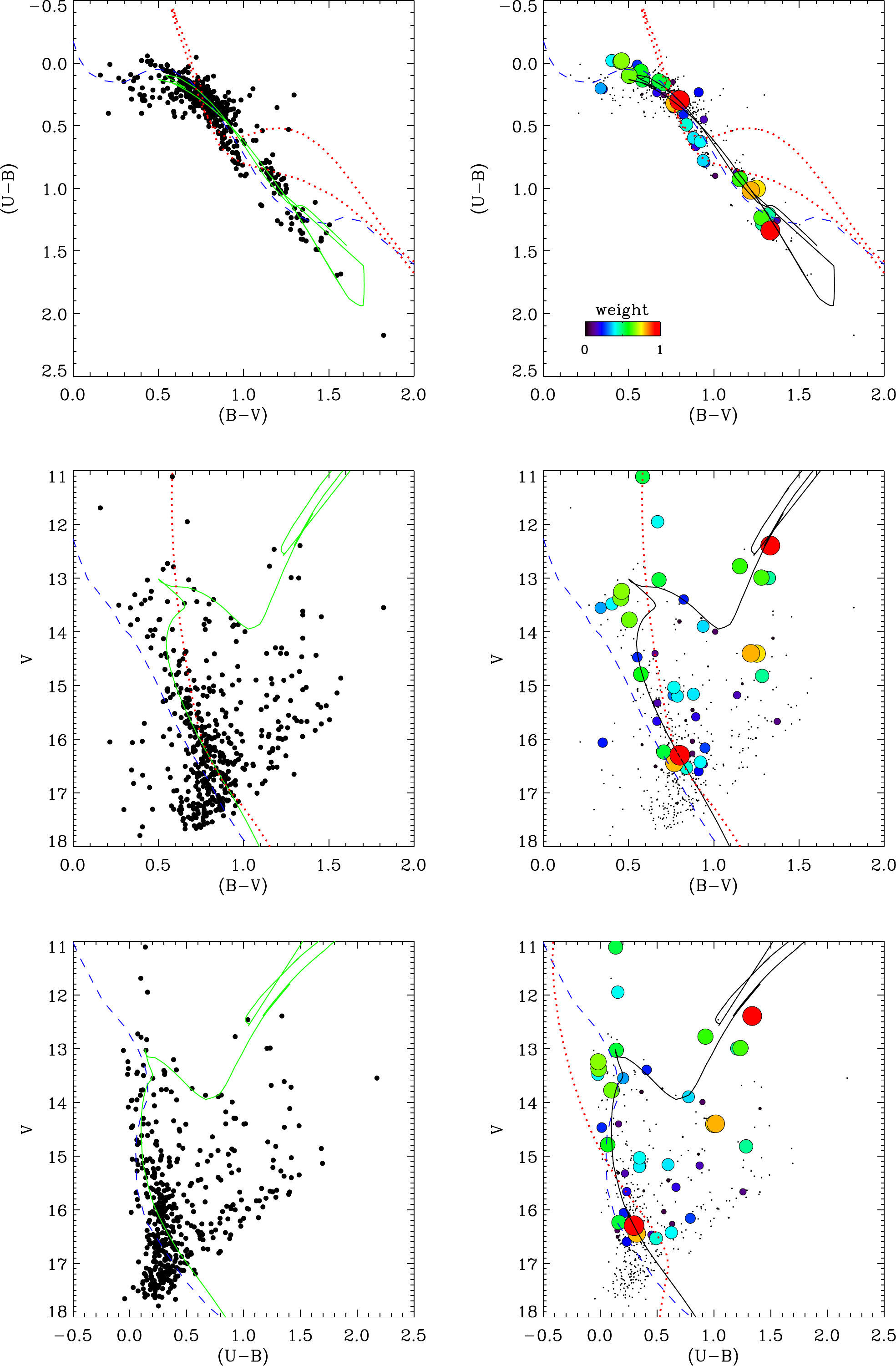]{Same as Fig. \ref{ngc2482}, but
for Ruprecht 154.\label{rup154}
}

\figcaption[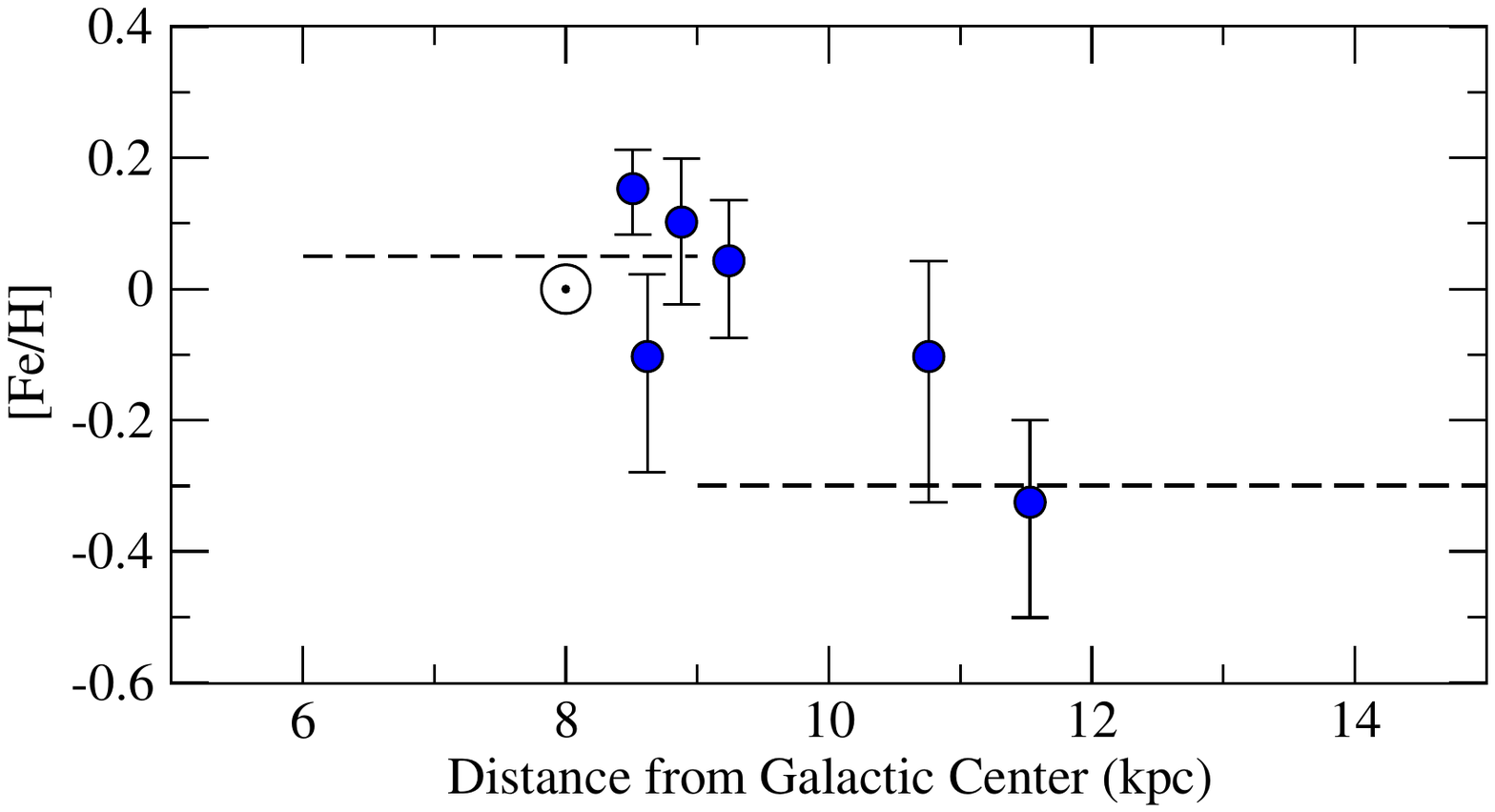]{Metallicity [Fe/H] of clusters vs. 
cluster distance from Galactic Center (in kpc).  The Sun's
distance is taken to be 8.0 kpc from the Galactic Center.
The dashed lines are based on \citet{Lep_etal11}.\label{metals}
}

\figcaption[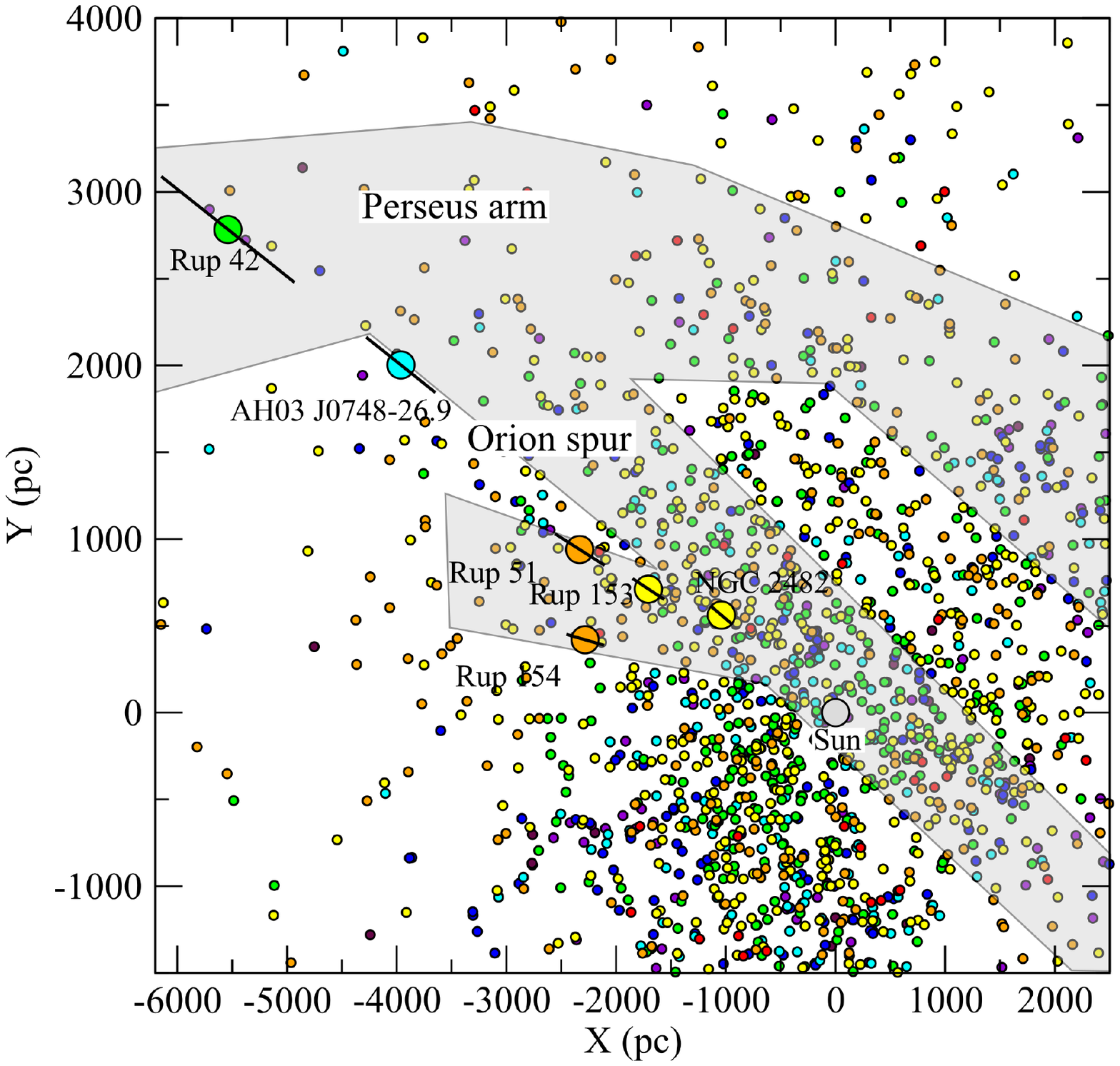]{Locations of clusters in the Galactic
X-Y plane.  Here we adopt the convention that
+X is in the direction of Galactic longitude 90\arcdeg.
+Y is in the direction of Galactic longitude 180\arcdeg.
Small colored dots: from \citet{Dia_etal02}.  Large colored dots:
data of this paper.  Large grey dot: position of the Sun.
The points are color coded by age.  The data are
binned by 1.0 in log $t$ (in years).
From the youngest clusters to the oldest the color coding is as
follows: blue (log $t$ from 6.0 to 7.0$^-$), 
green (7.0 to 8.0$^-$), yellow (8.0 to 9.0$^-$), and red (9.0 to 10.0).
The approximate boundaries of the Orion Spur and the Perseus arm are
from Fig. 16 of \citet{Chu_etal09}. \label{dias_cat}
}

\clearpage

\begin{figure}
\plotone{f1.pdf} {Fig. \ref{ngc2482_rgr}. 
}
\end{figure}

\begin{figure}
\epsscale{0.7}
\plotone{f2.pdf} {Fig. \ref{ngc2482}. 
}
\end{figure}

\begin{figure}
\plotone{f3.pdf} {Fig. \ref{finder}. 
}
\end{figure}

\begin{figure}
\plotone{f4.pdf} {Fig. \ref{pm_comp}. 
}
\end{figure}

\begin{figure}
\epsscale{0.7}
\plottwo{f5a.pdf}{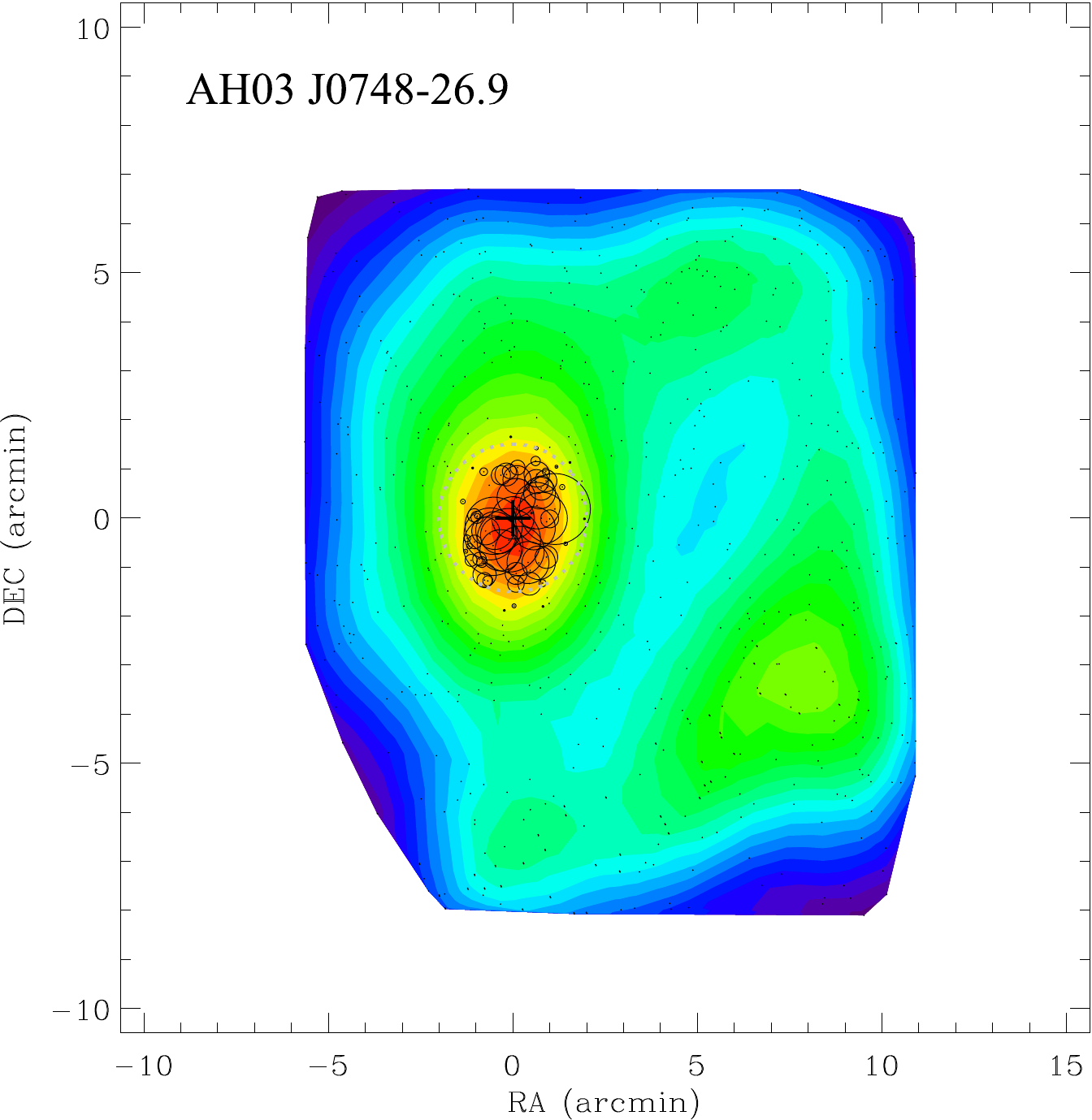}
\newline
\plottwo{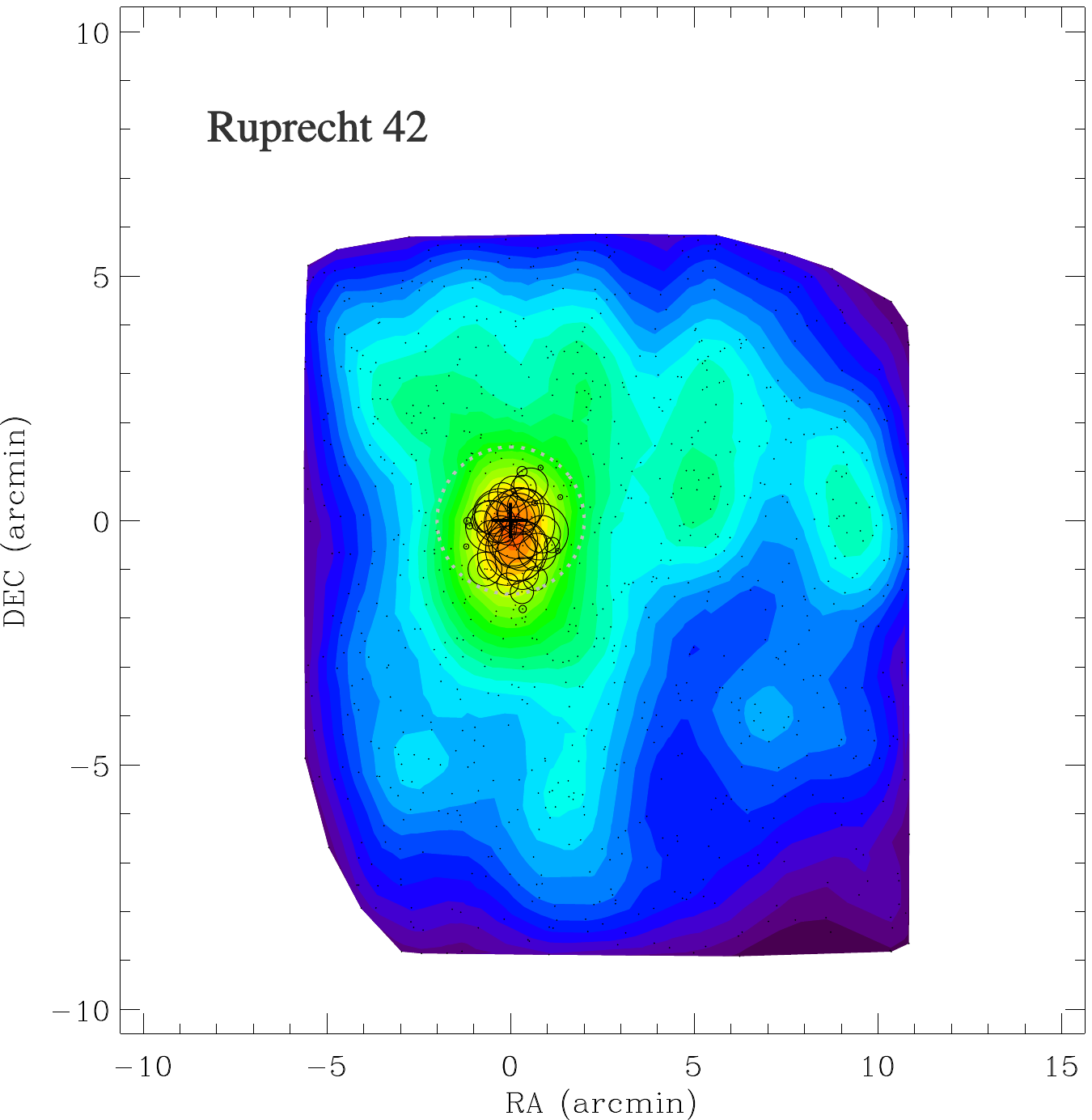}{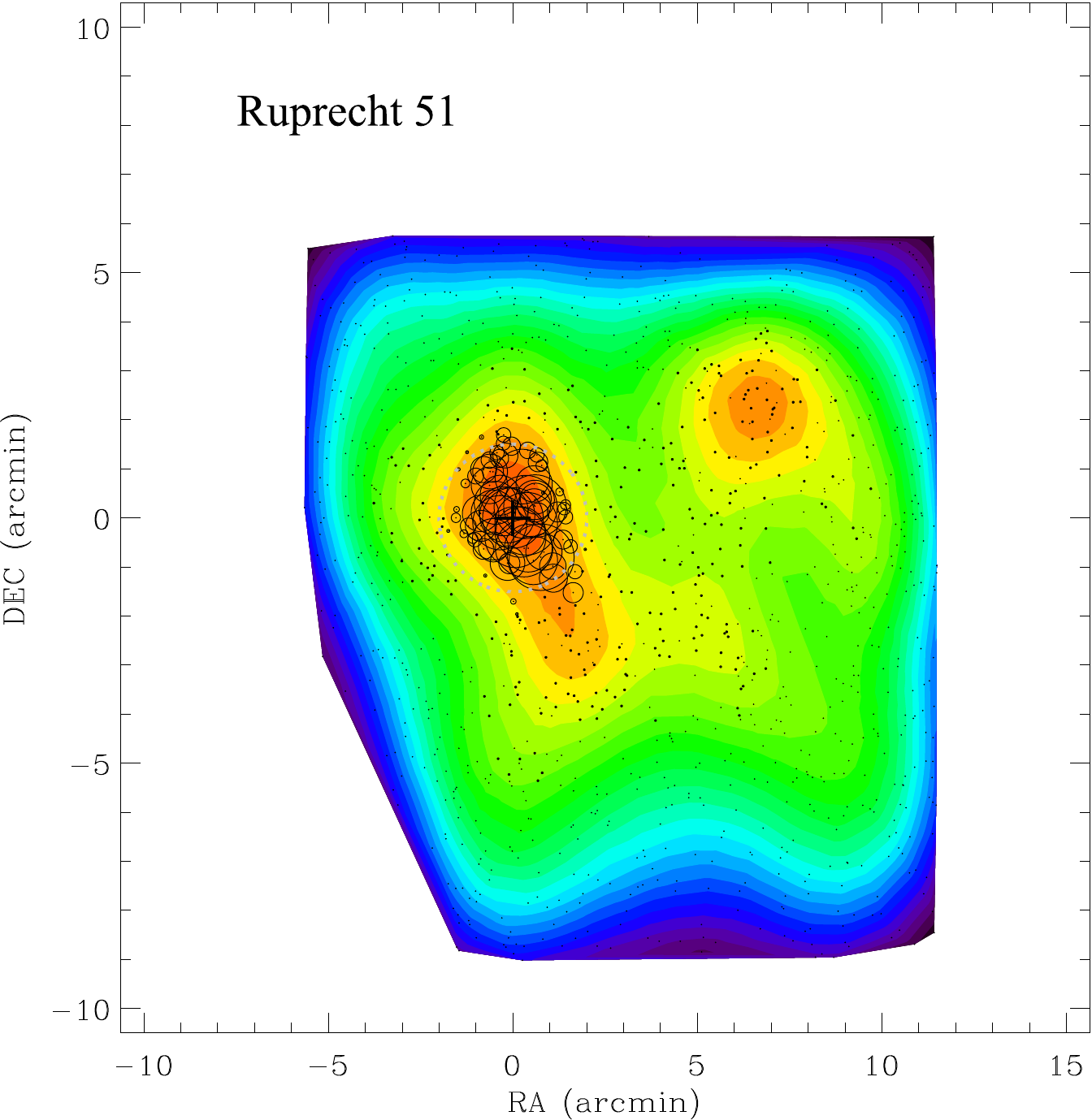}
\newline 
\plottwo{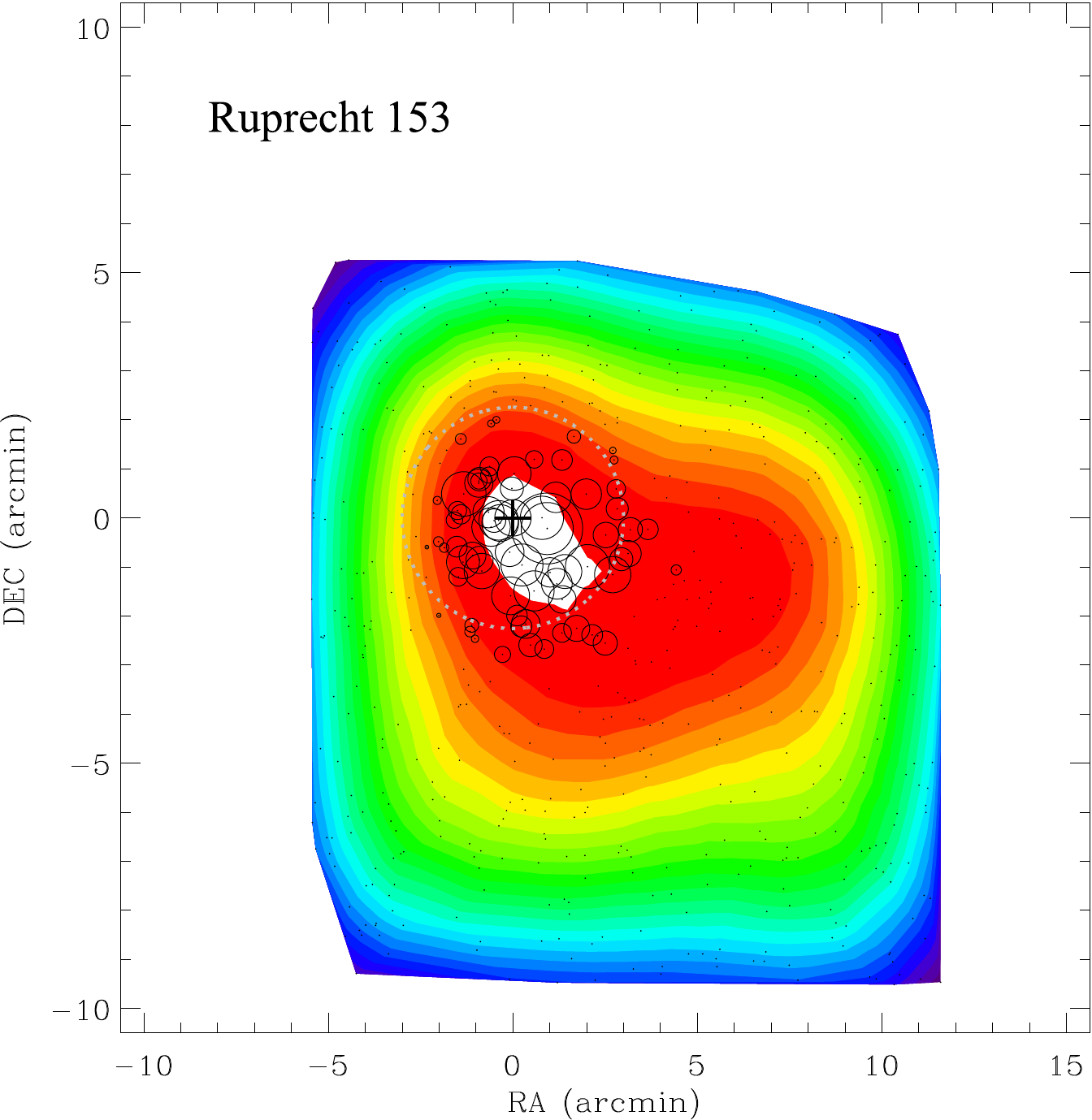}{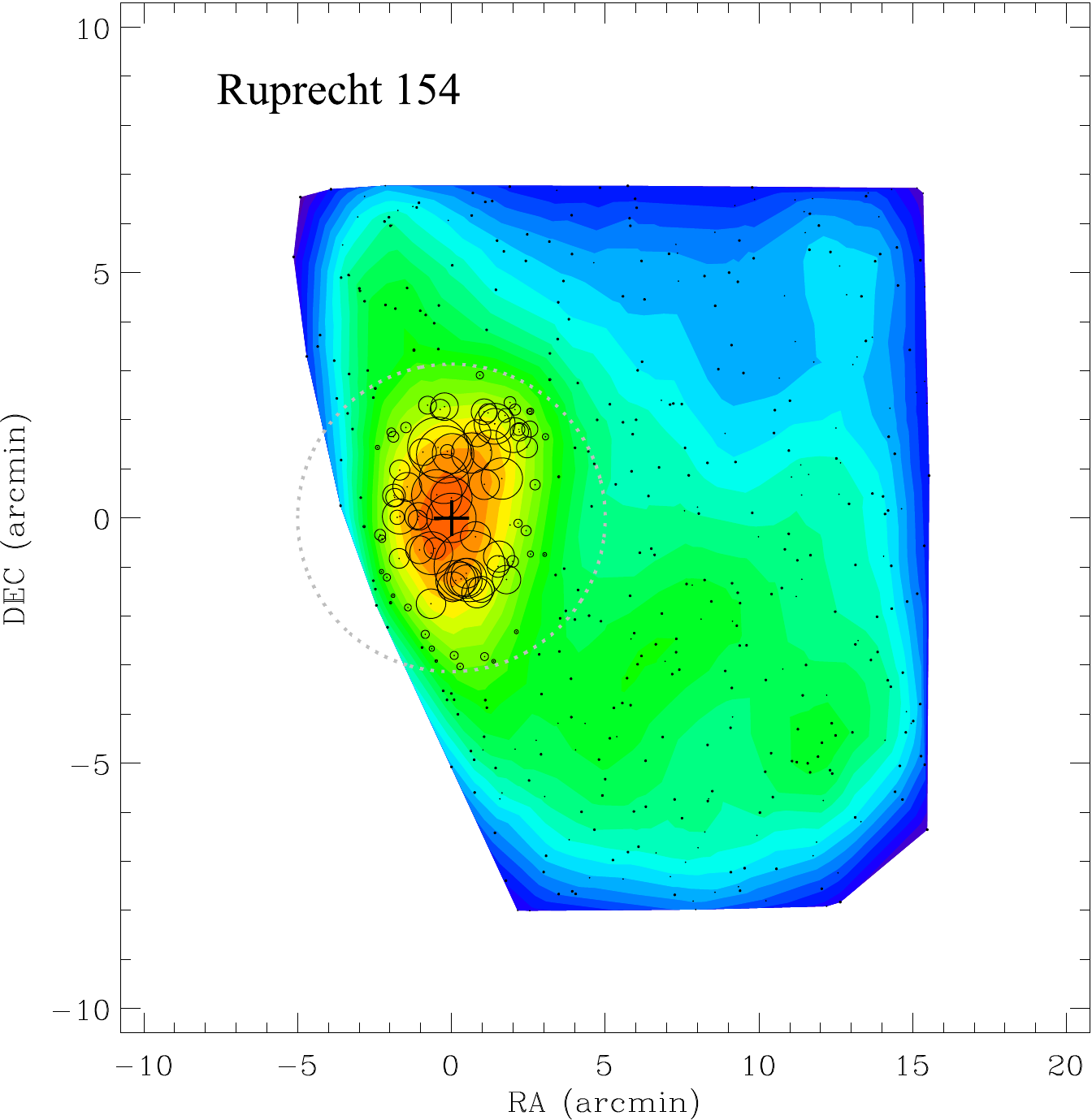} {Fig. \ref{density}. 
}
\end{figure}

\clearpage

\begin{figure}
\epsscale{0.7}
\plotone{f6.pdf} {Fig. \ref{rup154-hists}. 
}
\end{figure}

\begin{figure}
\epsscale{0.7}
\plotone{f7.pdf} {Fig. \ref{ah03}. 
}
\end{figure}

\begin{figure}
\epsscale{0.7}
\plotone{f8.pdf} {Fig. \ref{rup42}. 
}
\end{figure}

\begin{figure}
\epsscale{0.7}
\plotone{f9.pdf} {Fig. \ref{rup51}. 
}
\end{figure}

\begin{figure}
\epsscale{0.7}
\plotone{f10.pdf} {Fig. \ref{rup153}. 
}
\end{figure}

\begin{figure}
\epsscale{0.7}
\plotone{f11.pdf} {Fig. \ref{rup154}. 
}
\end{figure}
\epsscale{1.0}
\begin{figure}
\plotone{f12.pdf} {Fig. \ref{metals}. 
}
\end{figure}
\begin{figure}
\plotone{f13.pdf} {Fig. \ref{dias_cat}. 
}
\end{figure}

\end{document}